\documentclass[aps,twocolumn,showpacs,preprintnumbers,amsmath,amssymb,nofootinbib,superscriptaddress,showkeys]{revtex4}

\usepackage{hyperref}
\usepackage{float}
\usepackage{epsfig}
\usepackage{graphicx}
\usepackage{mathptmx}      
\usepackage{latexsym}
\usepackage{longtable}
\usepackage{dcolumn}

\begin{document}

\title{Statistical Error analysis of Nucleon-Nucleon phenomenological
  potentials}~\thanks{Supported by Spanish DGI (grant FIS2011-24149)
  and Junta de Andaluc{\'{\i}a} (grant FQM225).  R.N.P. is supported
  by a Mexican CONACYT grant.}

\author{R. Navarro P\'erez}\email{rnavarrop@ugr.es}
\affiliation{Departamento de F\'{\i}sica At\'omica, Molecular y
  Nuclear \\ and Instituto Carlos I de F{\'\i}sica Te\'orica y
Computacional \\ Universidad de Granada, E-18071 Granada, Spain.}
\author{J.E. Amaro}\email{amaro@ugr.es} \affiliation{Departamento de
  F\'{\i}sica At\'omica, Molecular y Nuclear \\ and Instituto Carlos I
  de F{\'\i}sica Te\'orica y Computacional \\ Universidad de Granada,
  E-18071 Granada, Spain.}  \author{E. Ruiz
  Arriola}\email{earriola@ugr.es} \affiliation{Departamento de
  F\'{\i}sica At\'omica, Molecular y Nuclear \\ and Instituto Carlos I
  de F{\'\i}sica Te\'orica y Computacional \\ Universidad de Granada,
  E-18071 Granada, Spain.}

\date{\today}

\begin{abstract} 
\rule{0ex}{3ex} Nucleon-Nucleon potentials are commonplace in nuclear
physics and are determined from a finite number of experimental data
with limited precision sampling the scattering process. We study the
statistical assumptions implicit in the standard least squares
$\chi^2$ fitting procedure and apply, along with more conventional
tests, a tail sensitive quantile-quantile test as a simple and
confident tool to verify the normality of residuals. We show that the
fulfillment of normality tests is linked to a judicious and consistent
selection of a nucleon-nucleon database. These considerations prove
crucial to a proper statistical error analysis and uncertainty
propagation. We illustrate these issues by analyzing about 8000
proton-proton and neutron-proton scattering published data. This
enables the construction of potentials meeting all statistical
requirements necessary for statistical uncertainty estimates in
nuclear structure calculations. 
\end{abstract}

\maketitle

\section{Introduction}

Nucleon-Nucleon potentials are the starting point for many nuclear
physics applications~\cite{Machleidt:1989tm}. Most of the current
information is obtained from np and pp scattering data below pion
production threshold and deuteron properties for which abundant
experimental data exist. The NN scattering amplitude reads  
\begin{eqnarray}
M 
&=& a
+ m (\mathbf{\sigma}_1\cdot\mathbf{n})(\mathbf{\sigma}_2\cdot\mathbf{n}) 
+ (g-h)(\mathbf{\sigma}_1\cdot\mathbf{m})(\mathbf{\sigma}_2\cdot\mathbf{m}) 
\nonumber \\ 
&+& 
(g+h)(\mathbf{\sigma}_1\cdot\mathbf{l})(\mathbf{\sigma}_2\cdot\mathbf{l})  
+ c (\mathbf{\sigma}_1+\mathbf{\sigma}_2)\cdot\mathbf{n}
\label{eq:wolfenstein} 
\end{eqnarray}
where $a,m,g,h,c$ depend on energy and angle, $\mathbf{\sigma}_1$ and
$\mathbf{\sigma}_2$ are the single-nucleon Pauli matrices,
$\mathbf{l}$, $\mathbf{m}$, $\mathbf{n}$ are three unitary orthogonal
vectors along the directions of $\mathbf{k}_f+\mathbf{k}_i$,
$\mathbf{k}_f-\mathbf{k}_i$ and $\mathbf{k}_i \wedge \mathbf{k}_f$, 
respectively,  and
($\mathbf{k}_f$, $\mathbf{k}_i$) are the final and initial relative
nucleon momenta. From these five complex energy and angle
dependent quantities 24 measurable cross-sections and polarization
asymmetries can be deduced~\cite{glockle1983quantum}. Inversely, a
complete set of experiments can be designed to reconstruct the
amplitude at a given energy~\cite{puzikov1957construction}. The finite
amount, precision and limited energy range of the data as well as the
many different observables calls for a standard statistical
$\chi^2$-fit
analysis~\cite{evans2004probability,eadie2006statistical}. This
approach is subjected to assumptions and applicability conditions that
can only be checked {\it a posteriori} in order to guarantee the
self-consistency of the analysis. Indeed, scattering experiments deal
with counting poissonian statistics and for moderately large number of
counts a normal distribution is expected.  Thus, one hopes that a
satisfactory theoretical description $O_i^{\rm th}$ can predict a set
of N independent observed 
data $O_i$ given an experimental uncertainty $\Delta O_i$ as
\begin{eqnarray}
O_i=O_i^{\rm th} + \xi_i \Delta O_i 
\label{eq:generator}
\end{eqnarray}
with $i=1, \dots, N$ and $\xi_i$ are independent random {\it normal}
variables with vanishing mean value $\langle \xi_i \rangle =0$ and
unit variance $\langle \xi_i \xi_j \rangle = \delta_{ij}$, implying
that $\langle O_i \rangle =O_i^{\rm th} $. Establishing the validity
of Eq.~(\ref{eq:generator}) is of utmost importance since it provides
a basis for the statistical interpretation of the error analysis. In
this work we will study to what extent this normality assumption
underlying the validity of the full $\chi^2$ approach is
justified. This will be done by looking at the statistical
distribution of the fit residuals of about 8000 np and pp published
scattering data collected since 1950. Using the normality test as a
necessary requirement we show that it is possible to fulfill
Eq.~(\ref{eq:generator}) with a high confidence level and high
statistics. Furthermore, we discuss the consequences and requirements
regarding the evaluation, design and statistical uncertainties of
phenomenological nuclear forces. We illustrate our points by
determining for the first time a smooth nuclear potential with {\it
  error bands} directly inferred from experiment. We hope that these
estimates will be useful for NN potential users interested in
quantifying a definite source of error in nuclear structure
calculations~\footnote{We note that in a Physical Review A
  editorial~\cite{PhysRevA.83.040001} the importance of including
  error estimates in papers involving theoretical evaluations has been
  stressed.}.

The history of $\chi^2$ statistical analyzes of NN scattering data
around pion production threshold started in the mid
fifties~\cite{Stapp:1956mz} (an account up to 1966 can be traced from
Ref.~\cite{arndt1966chi}). A modified $\chi^2$ method was
introduced~\cite{perring1963nucleon} in order to include data without
absolute normalization.  The steady increase along the years in the
number of scattering data and their precision generated mutually
incompatible data and hence a rejection criterion was
introduced~\cite{arndt1966determination,MacGregor:1968zzd,Arndt:1973ec}
allowing to discard inconsistent data. The upgrading of an ever
increasing consistent database poses the question of normality,
Eq.~(\ref{eq:generator}), of a large number of selected data. The
normality of the absolute value of residuals in pp scattering was
scrutinized and satisfactorily
fulfilled~\cite{Bergervoet:1988zz,Stoks:1992ja} as a necessary
consistency condition. The Nijmegen group made 20 years ago an important 
breakthrough by performing the very first Partial Wave Analysis (PWA) fit
with $\chi^2/{\rm dof} \sim 1$ and applying a $3\sigma$-rejection
criterion. This was possible after including em corrections, vacuum
polarization, magnetic moments interaction and a charge-dependent (CD)
One Pion Exchange (OPE) potential.  With this fixed database further
high quality potentials have been steadily
generated~\cite{Stoks:1994wp,Wiringa:1994wb,Machleidt:2000ge,Gross:2008ps}
and applied to nuclear structure calculations. However, high quality
potentials, i. e. those whose discrepancies with the data are
confidently attributable to statistical fluctuations in the
experimental data, have been built and used as if they were
error-less. As a natural consequence the computational accuracy to a
relative small percentage level has been a goal {\it per se} in the
solution of the few and many body problem regardless on the absolute
accuracy implied by the input of the calculation. While this sets high
standards on the numerical methods there is no {\it a priori} reason
to assume the computational accuracy reflects the realistic physical accuracy
 and, in fact, it would be extremely useful to determine
and identify the main source of uncertainty; one could thus tune the
remaining uncertainties to this less demanding level.

It should be noted that the $\chi^2$ fitting procedure, when applied
to limited upper energies, fixes most efficiently the long range piece
of the potential which is known to be mainly described by
One-Pion-Exchange (OPE) for distances $\, r \gtrsim 3 \, {\rm fm}$.
However, weaker constraints are put in the mid-range $r \sim 1.5-2.5
\, {\rm fm}$ region, which is precisely the relevant inter-particle
distance operating in the nuclear binding.  To date and to the best of
our knowledge, the estimation of errors in the nuclear force stemming
from the experimental scattering data uncertainties and its
consequences for nuclear structure calculations has not been seriously
confronted.  With this goal in mind we have upgraded the NN database
to include all published np and pp scattering data in the period
1950-2013, determining in passing the error in the
interaction~\cite{NavarroPerez:2012qf,Perez:2013za}. 

The present paper represents an effort towards filling this gap by
providing statistical error bands in the NN interaction based directly
on the experimental data uncertainties. In order to do so, the
specific form of the potential needs to be fixed~\footnote{This is
  also the case in the quantum mechanical inverse scattering problem,
  which has only unique solutions for specific assumptions on the form
  of the potential~\cite{chadan2011inverse} and with the additional
  requirement that some interpolation of scattering data at
  non-measured energies is needed. One needs then the information on
  the bound state energies and their residues in the scattering
  amplitude. We will likewise impose that the only bound state is the
  deuteron and reject fits with spurious bound states.}. As such, this
choice represents a certain bias and hence corresponds to a likely
source of systematic error. Based on the previous high quality fits
which achieved $\chi^2/\nu \lesssim 1
$~\cite{Stoks:1994wp,Wiringa:1994wb,Machleidt:2000ge,Gross:2008ps} we
have recently raised suspicions on the dominance of such errors with
intriguing consequences for the quantitative predictive power of
nuclear theory~\cite{NavarroPerez:2012vr,Perez:2012kt,Amaro:2013zka};
a rough estimate suggested that NN uncertainties propagate into an
unpleasantly large uncertainty of $\Delta B/A \sim 0.1-0.5 {\rm MeV}$,
a figure which has not yet been disputed by an alternative
estimate. In view of this surprising finding, there is a pressing need
to pin down the input uncertainties more accurately based on a variety
of sources~\footnote{There is a growing concern on the theoretical
  determination of nuclear masses from nuclear mean field models with
  uncertainty evaluation~\cite{Toivanen:2008im} (for a comprehensive
  discussion see e.g.~\cite{dudek2013predictive,Dobaczewski:2014jga})
  echoing the need for uncertainty estimates in a Physical Review A
  editorial~\cite{PhysRevA.83.040001} and the Saltelli-Funtowicz seven
  rules check-list~\cite{saltelli2013all}.}. This work faces the
evaluation of statistical errors after checking that the necessary
normality conditions of residuals and hence Eq.~(\ref{eq:generator})
are confidently fulfilled.  From this point of view, the present
investigation represents an initial step, postponing a more complete
discussion on systematic uncertainties for a future investigation.

The PWA analysis carried out previously by
us~\cite{NavarroPerez:2012vr,Perez:2012kt,Amaro:2013zka}; was
computationally inexpensive due to the use of the simplified
$\delta$-shell potential suggested many years ago by
Avil\'es~\cite{Aviles:1973ee}. This form of potential effectively
coarse grains the interaction and drastically reduces the number of
integration points in the numerical solution of the Schr\"odinger
equation (see e.g.~\cite{NavarroPerez:2011fm}).  However, it is not
directly applicable to some of the many numerical methods available on
the market to solve the few and many body problem where a smooth
potential is required. Therefore, we will analyze the $3\sigma$-self
consistent database in terms of a more conventional potential form
containing the same 21-operators extending the AV18 as we did
in\cite{NavarroPerez:2012vr,Perez:2012kt,Amaro:2013zka}.  Testing for
normality of residuals within a given confidence level for a {\it
  phenomenological potential} is an issue of direct
significance to any statistical error analysis and
propagation. Actually, we will show that for the fitted observables to
the $3\sigma$ self consistent experimental data base $O_i^{\rm exp}$,
with quoted uncertainty $\Delta O_i^{\rm exp}$, $i=1, \dots,
N=6713$ (total number of pp and np scattering data), our theoretical
fits indeed satisfy that the residuals
\begin{eqnarray}
R_i=\frac{O_i^{\rm exp} - O_i^{\rm th}}{\Delta O_i^{\rm exp}} 
\label{eq:residuals}
\end{eqnarray}
follow a normal distribution within a large confidence level. In order
to establish this we will use a variety of classical statistical
tests~\cite{evans2004probability,eadie2006statistical}, such as the
Pearson, Kolmogorov-Smirnov (KS), the moments method (MM) and, most
importantly, the recently proposed Tail-Sensitive (TS)
quantile-quantile test with confidence bands~\cite{Aldor2013}.  By
comparing with others the TS test turns out to be the most demanding
one concerning the confidence bands.  Surprisingly, normality tests
have only seldom been applied within the present context, so our
presentation is intended to be at a comprehensive level. A notable
exception is given in Refs.~\cite{Bergervoet:1988zz,Stoks:1992ja}
where the moments method in a pp analysis up to $T_{\rm LAB}=30$ and
350 MeV, is used for $N=389$ and 1787 data, respectively, to test that
the squared residuals $R_i^2$ in Eq.~(\ref{eq:residuals}) follow a
$\chi^2$-distribution with one degree of freedom. Note, that this is
insensitive to the sign of $R_i$ and thus blind to asymmetries in a
normal distribution. Here, we test normality of $R_i$ for a total of
$N=6713$ np and pp data up to $T_{\rm LAB}=350$ MeV.

The paper is organized as follows. In Section~\ref{sec:statistics} we
review the assumptions, the rejecting and fitting process used in our
previous works to build the $3\sigma$ self-consistent database, and
expose the main motivation to carry out a normality test of the fit
residuals. In Section~\ref{sec:normal} we review some of the classical
normality tests and a recently proposed tail-sensitive test, which we
apply comparatively to the complete as well as the $3\sigma$
self-consistent database, providing a {\it raison d'\^etre} for the
rejection procedure. After that, in Section~\ref{sec:OPEgauss} we
analyze a fit of a potential whose short distance
contribution is constructed by a sum of Gaussian functions, with particular
attention to the error bar estimation, a viable task since the
residuals pass satisfactorily the normality test. Finally in
Section~\ref{sec:conclusions} we come to our conclusions and provide
some outlook for further work.

\section{Statistical Considerations}
\label{sec:statistics}

There is a plethora of references on data and error analysis (see
e.g.~\cite{evans2004probability,eadie2006statistical}). We will review
the fitting approach in such a  way that our points can be more easily
stated for the general reader.

\subsection{Data uncertainties}

Scattering experiments are based on counting Poissonian statistics and
for moderately large number of counts a normal distribution sets in.
In what follows $O_i$ will represent some scattering observable.  For
a set of $N$ independent measurements of different scattering
observables $O_i^{\rm exp}$ experimentalists quote an {\it estimate}
of the uncertainty $ \Delta O_i^{\rm exp}$ so that the {\it true}
value $O_i^{\rm true}$ is contained in the interval $O_i^{\rm exp} \pm
\Delta O_i^{\rm exp} $ with a $68\%$ confidence level. In what follows
we assume for simplicity that there are no sources of systematic
errors. Actually, when only the pair $(O_i^{\rm exp},\Delta O_i^{\rm
  exp}) $ is provided without specifying the distribution we will
assume an underlying normal distribution~\footnote{This may not be the
  most efficient unbiased estimator (see
  e.g.~\cite{evans2004probability,eadie2006statistical} for a more
  thorough discussion). Quite generally, the theory for the noise on
  the specific measurement would involve many considerations on the
  different experimental setups. In our case the many different
  experiments makes such an approach unfeasible. There is a
  possibility that some isolated systematic errors in particular
  experiments are randomized when considered globally. However, the
  larger the set the more stringent the corresponding statistical
  normality test. From this point of view the verification of the
  normality assumption underlying Eq.~(\ref{eq:generator}) proves
  highly non-trivial.}, so that
\begin{eqnarray}
P ( O_i^{\rm exp} ) = 
\frac{\exp\left[-\frac12 \left(
  \frac{O_i^{\rm true}-O_i^{\rm exp}}{\Delta O^{\rm exp}} \right)^2 \right]}{\sqrt{2\pi} \Delta O_i^{\rm exp}} 
\end{eqnarray}
is the probability density of finding measurement $O_i^{\rm exp}$.

\subsection{Data modeling}
\label{sec:modeling}

The problem of data modeling is to find a theoretical description
characterized by some parameters $F_i (\lambda_1, \dots , \lambda_P)$
which contain the true value $O_i^{\rm true} = F_i (\lambda_1^{\rm
  true}, \dots , \lambda_P^{\rm true})$ with a given confidence level
characterized by a bounded p-dimensional manifold in the space of
parameters $(\lambda_1, \dots , \lambda_P)$. For a normal distribution
the probability of finding any of the ({\it independent} )
measurements $O_i^{\rm exp}$, assuming that $(\lambda_1, \dots , \lambda_P)$ are
the true parameters, is given by
\begin{eqnarray}
P ( O_i^{\rm exp} | \lambda_1 \dots \lambda_P ) = \frac{\exp\left[-\frac12 \left(
  \frac{F_i(\lambda_1 , \dots, \lambda_P)-O_i^{\rm exp}}{\Delta O^{\rm exp}} \right)^2 \right]}{\sqrt{2\pi} \Delta O_i^{\rm exp}} 
\end{eqnarray}
Thus the joined probability density is 
\begin{eqnarray}
P(O_1^{\rm exp} \dots O_N^{\rm exp} | \lambda_1 \dots \lambda_P ) &=& \prod_{i=1}^N P(O_i^{\rm exp} |
\lambda_1 \dots \lambda_P ) \nonumber \\ 
&=& C_N e^{-\chi^2 (\lambda_1, \dots, \lambda_P)/2}
\end{eqnarray}
where $1/C_N = \prod_{i=1}^N (\sqrt{2\pi} \Delta O_i^{\rm exp})$.  In
such a case the maximum likelihood
method~\cite{evans2004probability,eadie2006statistical} corresponds to
take the $\chi^2$ as a figure of merit given by
\begin{eqnarray}
\chi^2(\lambda_1, \dots, \lambda_P)
= \sum_{i=1}^{N} \left( \frac{O_i^{\rm exp}-F_i(\lambda_1, \dots,
  \lambda_P)}{\Delta O_i^{\rm exp}}\right)^2
\end{eqnarray}
and look for the minimum in the fitting parameters $(\lambda_1, \dots,
\lambda_P)$, 
\begin{eqnarray}
\chi^2_{\rm min}= \min_{\lambda_i} \chi^2 (\lambda_{1}, \dots ,
\lambda_{P}) = \chi^2 (\lambda_{1,0}, \dots , \lambda_{P,0})
\end{eqnarray}
Our theoretical estimate of $O_i^{\rm true}$ after the fit is given by 
\begin{eqnarray}
O_i^{\rm th}=F_i(\lambda_{1,0}, \dots,  \lambda_{P,0}).
\end{eqnarray}
Expanding around the minimum one has  
\begin{equation}
\chi^2 = \chi^2_{\rm min}+ \sum_{ij=1}^{P} (\lambda_i-\lambda_{i,0})
(\lambda_j-\lambda_{j,0}) {\cal E}^{-1}_{ij} + \cdots
\end{equation}
where the $P\times P$ 
error matrix is defined as the inverse of the Hessian matrix
evaluated at the minimum
\begin{equation}
 {\cal E}^{-1}_{ij}= \frac12 
\frac{\partial^2 \chi^ 2}{\partial \lambda_i
  \partial \lambda_j}(\lambda_{1,0},\ldots,\lambda_{P,0})
\end{equation}
Finally, the correlation matrix between two fitting parameters
$\lambda_i$ and $\lambda_j$ is given by
\begin{equation}
 {\cal C}_{ij}=  \frac{{\cal E}_{ij}}{\sqrt{{\cal E}_{ii}{\cal E}_{jj}}}
\label{eq:correlation}
\end{equation}

We compute the error of the parameter $\lambda_i$ as
\begin{equation}
\Delta\lambda_i \equiv \sqrt{\cal E}_{ii} .
\end{equation}
Error propagation of an observable $G=G(\lambda_1,\ldots,\lambda_P)$ 
is computed as
\begin{equation}
(\Delta G)^2 = \sum_{ij} 
\frac{\partial G }{\partial\lambda_i}
\frac{\partial G}{\partial\lambda_j}  \Big|_{\lambda_k=\lambda_{k,0}}
{\cal E}_{ij}.
\label{eq:error-prop}
\end{equation}
The resulting residuals of the fit are defined as 
\begin{equation}
 R_i = \frac{O_i^{\rm exp}-F_i (\lambda_{1,0} , \dots,
   \lambda_{P,0})}{\Delta O_i^{\rm exp}}, \ \ i = 1,...,N.
\end{equation}
Assuming normality of residuals is now crucial for an statistical
interpretation of the confidence level, since then $\sum_i R_i^2$
follows a $\chi^2$-distribution. One useful application of the
previous result is that we can replicate the experimental data by
using Eq.~(\ref{eq:generator}) and  in such a case $\langle
\chi^2 \rangle = N$. For a large number of data $N$ with
$P$-parameters one has, with a $1\sigma$ or $68\%$ confidence level,
the mean value and most likely value nearly coincide, so that one has
$ \langle \chi^2_{\rm min} \rangle = N-P $ and thus as a random
variable we have
\begin{eqnarray}
\frac{\chi^2_{\rm min}}{\nu} \equiv \frac{\sum_i \xi_i^2}\nu = 1 \pm
\sqrt{\frac{2}\nu}
 \end{eqnarray}
where $\nu=N-P$ is the number of degrees of freedom. The goodness
of fit is defined in terms of this confidence interval. However, the
$\chi^2$-test has a sign ambiguity for every single residual given
that $R_i \to - R_i$ is a symmetry of the test. From this point of
view the verification of normality is a more demanding
requirement~\footnote{One can easily see that for a set of normally
  distributed data $R_i$, while $|R_i|$ does not follow that
  distribution, $|R_i|^2=R_i^2$ would pass a $\chi^2$-test.}.

Thus a necessary condition for a least-squares fit with meaningful
results is the residuals to follow a normal distribution with mean
zero and variance one, i.e. $R_i \sim N(0,1)$. It should be noted that
a model for the noise need not be normal, but it must be a {\it known}
distribution $P(z)$ such that the residuals $R_i$ do indeed follow
{\it such} distribution~\footnote{In this case the merit figure to
  minimize would be
$$
S(\lambda_1, \dots , \lambda_P) =-\sum_i \log P \left( \frac{O_i^{\rm exp} - F_i (\lambda_1, \dots, \lambda_P)}{\Delta O_i^{\rm
    exp}}\right)$$ For instance in Ref.~\cite{Matsinos:1997pb}, dealing with
  $\pi N$ scattering a Lorentz distribution arose as a self consistent
  assumption.}.

\subsection{Data selection}

The first and most relevant problem one has to confront in the
phenomenological approach to the Nucleon-Nucleon interaction is that
the data base is not consistent; there appear to be incompatible
measurements. This may not necessarily mean genuinely wrong
experiments, but rather unrealistic error estimates or an incorrect
interpretation of the quoted error as a purely statistical
uncertainty~\footnote{Indeed any measurement could become right
  provided a sufficiently large or conservative error is
  quoted.}. Note that the main purpose of a fit is to estimate the
true values of certain parameters with a given and admissible
confidence level.  Therefore one has to make a decision on which are
the subset of data which will finally be used to determine the NN
potential. However, once the choice has been made the requirement of
having normal residuals, Eq.~(\ref{eq:residuals}), must be checked if
error estimates on the fitting parameters are truly based on a random
distribution.

The situation we encounter in practice is of a large number of data
$\sim 8000$ vs the small number of potential parameters $\sim 40$
which are expected to successfully account for the description of the
data~\cite{Perez:2013cza}. Thus, naively there seems to be a large
redundancy in the database. However, there is a crucial issue on what
errors have been quoted by the experimentalists. If a conservative
estimate of the error is made, there is a risk of making the
experiment useless, from the point of view that any other experiment
in a similar kinematical region will dominate the
analysis~\footnote{See e.g. the recommendations of the Guide to the
  Expression of Uncertainty in Measurement by the
  BIPM~\cite{bipm1995oiml} where (often generously) {\it conservative}
  error estimates are undesirable, while {\it realistic} error
  estimates are preferable. Of course, {\it optimal} error estimates
  could only arise when there is a competition between independent
  measurements and a bonus for accuracy.}. If, on the contrary, errors
are daringly too small, they may generate a large penalty as compared
to the rest of the database. This viewpoint seems to favor more
accurate measurements whenever they are compatible but less accurate
ones when some measurements appear as incompatible with the rest. In
addition, there may be an abundance bias, i.e. too many accurate
measurements in some specific kinematical region and a lack of
measurements in another regions. Thus, the working assumption in order
to start any constructive analysis is that {\it most} data have {\it
  realistic} quoted errors, and that those experiments with
unrealistically too small or too large errors can be discerned from
the bulk with appropriate statistical tools.  This means that these
unrealistic uncertainties can be used to reject the corresponding
data~\footnote{From this point of view, the small and the large errors
  are not symmetric; the small $\chi^2$ (conservative errors) indicate
  that the fitting parameters are indifferent to these data, whereas
  the large $\chi^ 2$ (daring errors) indicate an inconsistency with
  the rest of the data.}.  If a consistent and maximal database is
obtained by an iterative application of a rejection criterium, the
discrepancy between theory and data has to obey a statistical
distribution, see Eq.~(\ref{eq:generator}).

\begin{figure*}
\centering
\includegraphics[width=\linewidth]{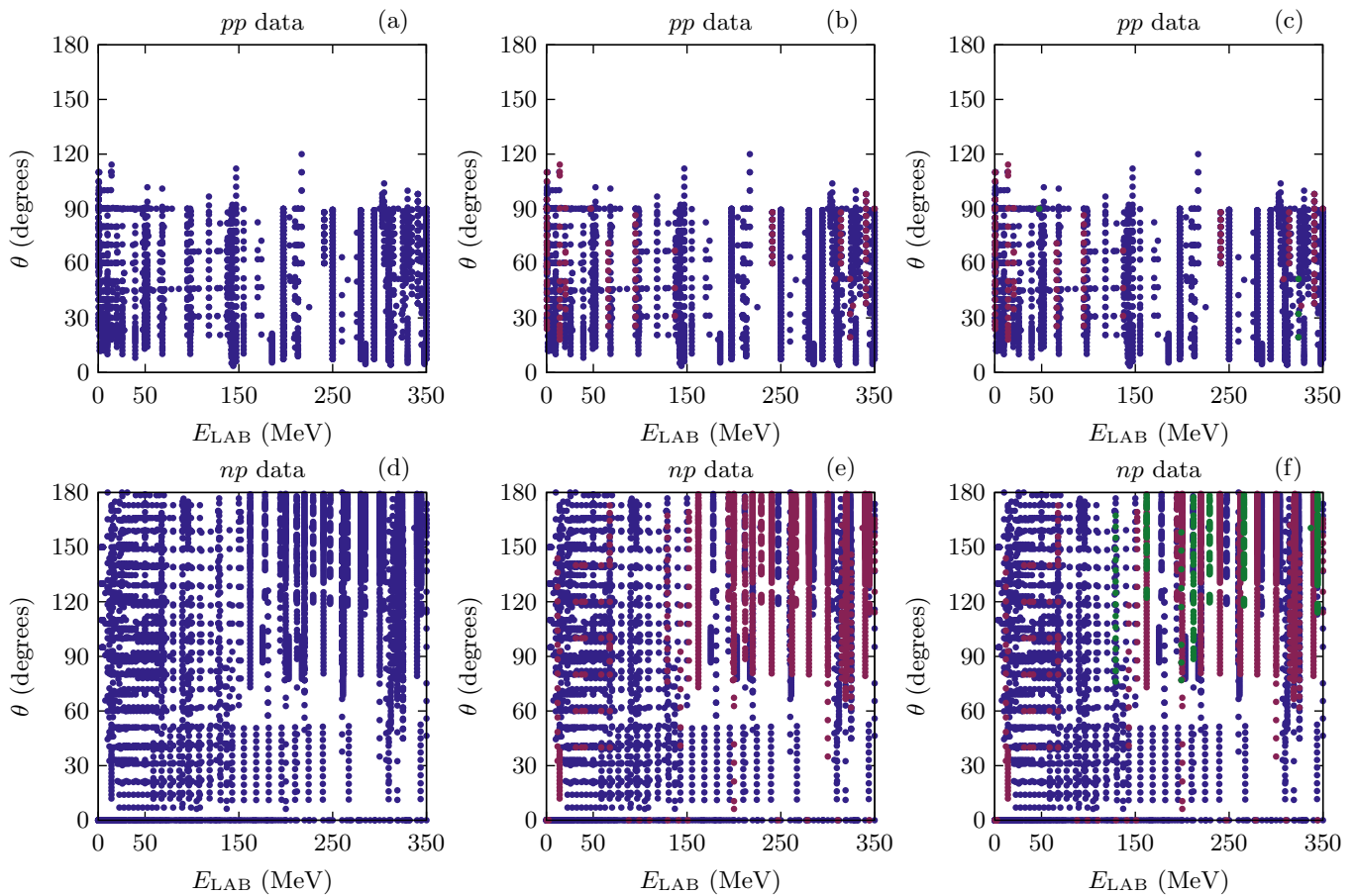}
\caption{(Color online) Abundance plots for pp (top panel) and np (bottom panel)
  scattering data. Full data base (left panel). Standard $3\sigma$
  criterion (middle panel). Self-consistent $3\sigma$ criterion (right
  panel). We show accepted data (blue), rejected data (red) and
  recovered data (green).  }
\label{fig:NN-abundance}       
\end{figure*}

\subsection{Data representation}

For two given data with exactly the same kinematical conditions, i.e.,
same observable, scattering angle and energy, the decision on whether
they are compatible or not may be easily made by looking at
non-overlapping error bands~\footnote{For several measurements the
  Birge test~\cite{birge1932calculation} is the appropriate tool. The
  classical and Bayesian interpretation of this test has been
  discussed recently~\cite{kacker2008classical}.}. This is frequently
not the case; one has instead a set of neighboring data in the
$(\theta,E)$ plane for a given observable or different observables at
the same $(\theta,E)$ point.  The situation is depicted in
Fig.~\ref{fig:NN-abundance} (left panels) where every point represents
a single pp or np measurement (for an illustrative plot on the
situation by 1983 up to $1{\rm GeV}$ see \cite{Arndt:1982ep}). The
total number of 8124 fitting data includes 7709 experimental
measurements and 415 normalizations provided by the
experimentalists. Thus, the decision intertwines all available data
and observables. As a consequence the comparison requires a certain
extrapolation, which is viable under a smoothness assumption of the
energy dependence of the partial wave scattering
amplitude. Fortunately, the meson exchange picture foresees a well
defined analytical branch cut structure in the complex energy plane
which is determined solely from the long distance properties of the
interaction.  A rather efficient way to incorporate this desirable
features from the start is by using a quantum mechanical potential.
More specifically, if one has $n\pi$ exchange then at long distances
$V(r) \sim e^{-n m_\pi r}$ guarantees the appearance of a left hand
branch cut at C.M. momentum $p= i m_\pi n/2$. Using this meson
exchange picture at long distances the data world can be mapped onto
a, hopefully complete, set of fitting parameters.

In order to analyze this in more detail we assume, as we did in
Refs.~\cite{NavarroPerez:2012vr,Perez:2012kt,Amaro:2013zka}, that the
NN interaction interaction can be decomposed as
\begin{eqnarray}
   V(\vec r) = V_{\rm short} (r) \theta(r_c-r)+ V_{\rm long} (r) \theta(r-r_c),
\label{eq:potential}
\end{eqnarray}
where  the short component can be written as 
\begin{eqnarray}
   V_{\rm short}(\vec r) = \sum_{n=1}^{21} \hat O_n \left[\sum_{i=1}^N V_{i,n} F_{i,n} (r)\right] 
\label{eq:potential-short}
\end{eqnarray}
where $ \hat O_n$ are the set of operators in the extended AV18
basis~\cite{Wiringa:1994wb,NavarroPerez:2012vr,Perez:2012kt,Amaro:2013zka},
$V_{i,n}$ are unknown coefficients to be determined from data and
$F_{i,n}(r) $ are some given radial functions.  $V_{\rm long}(\vec r)$
contains a Charge-Dependent (CD) One pion exchange (OPE) (with a
common
$f^2=0.075$~\cite{NavarroPerez:2012vr,Perez:2012kt,Amaro:2013zka}) and
electromagnetic (EM) corrections which are kept fixed throughout. This
corresponds to
\begin{eqnarray}
V_{\rm long}(\vec r) = V_{\rm OPE}(\vec r) + V_{\rm em}(\vec r) \, . 
\end{eqnarray}
Although the form of the complete potential is expressed in the
operator basis the statistical analysis is carried out more
effectively in terms of some low and independent partial waves
contributions to the potential from which all other higher partial
waves are consistently deduced (see
Ref.~\cite{Perez:2013mwa,Perez:2013jpa}).

\subsection{Fitting data}

In our previous PWA we used the delta-shell interaction already
proposed by Avil\'es~\cite{Aviles:1973ee} and which proved extremely
convenient for fast minimization and error evaluation~\footnote{We use
  the Levenberg-Marquardt method where an approximation to the Hessian
  is computed explicitly~\cite{press2007numerical} which we keep
  throughout.} and corresponds to the choice
\begin{eqnarray}
F_{i,n}(r) = \Delta r_i \delta(r-r_i) 
\end{eqnarray}
where $r_i \le r_c$ are a discrete set of radii and $\Delta r_i =
r_{i+1}-r_i $. The minimal resolution $\Delta r_{\rm min}$ is
determined by the shortest de Broglie wavelength corresponding to pion
production threshold which we estimate as $\Delta r_{\pi} \sim 0.6
{\rm fm}$~\cite{NavarroPerez:2011fm,Perez:2013cza} so that the needed
number of parameters can be estimated {\it a priori}. Obviously, if
$\Delta r_{\rm min} \ll \Delta r_\pi $ the number of parameters
increases but also the correlations among the different fitting
coefficients, $V_{i,n}$, so that some parameters become redundant or
an over-complete representation of the data, and the $\chi^2$ value
will not decrease substantially. In the opposite situation $\Delta
r_{\rm min} \gg \Delta r_\pi $ the coefficients $V_{i,n}$ do not
represent the database and hence are incomplete. Our fit with an
uniform $\Delta r \equiv \Delta r_\pi $ was satisfactory, as expected.

\subsection{The $3\sigma$ self-consistent database}

After the fitting process we get the desired $3\sigma$ self-consistent
database using the idea proposed by Gross and
Stadler~\cite{Gross:2008ps} and worked at full length in our previous
work~\cite{Perez:2013jpa}.  This allows to rescue data which would
otherwise have been discarded using the standard $3\sigma$ criterion
contemplated in all previous analyzes
~\cite{Stoks:1993tb,Stoks:1994wp,Wiringa:1994wb,Machleidt:2000ge,Gross:2008ps}. The
situation is illustrated in Fig.~\ref{fig:NN-abundance} (middle and
right panels).

By using the rejection criterion at the $3\sigma$ level we cut-off the
long tails and as a result a fair comparison could in principle be
made to this truncated gauss distribution. The Nijmegen group found
that the moments method test (see below for more details) largely
improved by using this truncated
distribution~\cite{Bergervoet:1988zz}.  It should be reminded,
however, that the rejection criterion is applied to groups of
datasets, and not to individual measurements, and in this way gets
coupled with the floating of normalization.  One could possibly
improve on this by trying to determine individual outliers in a
self-consistent way which could make a more flexible data selection.
Preliminary runs show that the number of iterations grows and the
convergence may be slowed down or non-converging by marginal decisions
with some individual data flowing in and out the acceptance
domain. Note also, that rejection may also occur because data are
themselves non normal or the disentanglement between statistical and
systematic errors was not explicitly exploited. In both cases these
data are useless to propagate uncertainties invoking the standard
statistical interpretation, see Eq.~(\ref{eq:error-prop}).

\subsection{Distribution of residuals}

\begin{figure}[hpt]
\begin{center}
\epsfig{figure=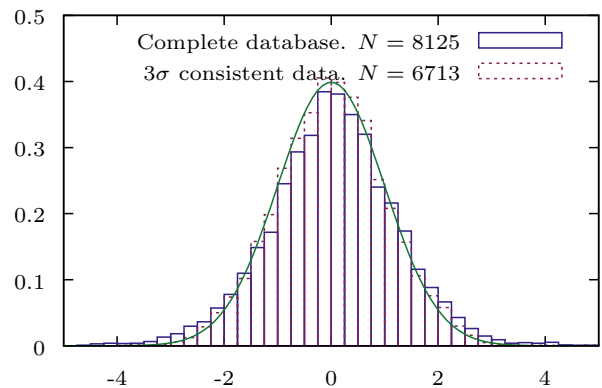,width=0.9\linewidth}  
\end{center}
\caption{(Color online) Normalized histogram of the resulting
  residuals after fitting the potential parameters to the complete
  $pp$ and $np$ database (blue boxes with solid borders) and to the $3
  \sigma$ consistent database (red boxes with dashed borders). The
  $N(0,1)$ standard normal probability distribution function (green
  solid line) is plotted for comparison.}
\label{Fig:Histograms}
\end{figure}

In Fig.~\ref{Fig:Histograms} we present the resulting residuals,
Eq.~(\ref{eq:residuals}), in a normalized histogram for illustration
purposes, in the case of the original full database and the
$3\sigma$-consistent database, and compare them with a normal
distribution function with the binning resolution $\Delta R =
0.2$. The complete database histogram shows an asymmetry or skewness
as well as higher tails and clearly deviates from the normal
distribution, meanwhile the $3 \sigma$ consistent database residuals
exhibits a closer agreement with the Gaussian distribution. Note that
this perception from the figure is somewhat depending on eyeball
comparison of the three situations. We will discuss more preferable
tests in the next section which are independent on this binning
choice.

\begin{table*}
 \caption{\label{tab:Cmoments} Standardized moments $\mu_r'$ of the
   residuals obtained by fitting the complete database with the
   delta-shell potential and $3\sigma$ consistent database with the
   OPE-delta-shell, $\chi$TPE-delta-shell and OPE-Gaussian
   potentials. The expected values for a normal distributions are
   included $\pm$ $1 \sigma$ confidence level of a Monte Carlo
   simulation with $5000$ random samples of size $N$}
 \begin{ruledtabular}
 \begin{tabular*}{\columnwidth}{@{\extracolsep{\fill}} c D{.}{.}{6.3} D{.}{.}{3.3} D{.}{.}{6.3} D{.}{.}{2.3} D{.}{.}{5.3} D{.}{.}{3.3}D{.}{.}{5.3} D{.}{.}{3.3}}
       & \multicolumn{2}{c}{Complete data base} & \multicolumn{2}{c}{$3\sigma$ OPE-Delta-Shell} & \multicolumn{2}{c}{$3\sigma$ $\chi$TPE-Delta-Shell} & \multicolumn{2}{c}{$3\sigma$ OPE-Gaussian} \\
       & \multicolumn{2}{c}{$N=8125$} & \multicolumn{2}{c}{$N=6713$} & \multicolumn{2}{c}{$N=6712$} & \multicolumn{2}{c}{$N=6711$} \\
   $r$ & \multicolumn{1}{c}{Expected} & \multicolumn{1}{c}{Empirical} 
       & \multicolumn{1}{c}{Expected} & \multicolumn{1}{c}{Empirical} 
       & \multicolumn{1}{c}{Expected} & \multicolumn{1}{c}{Empirical} 
       & \multicolumn{1}{c}{Expected} & \multicolumn{1}{c}{Empirical} \\
  \hline
    3 &   0 \pm \hphantom{0}0.027 &  -0.176 &   0 \pm \hphantom{0}0.030 &  0.007 &   0 \pm \hphantom{0}0.030 &  -0.011 &   0 \pm \hphantom{0}0.030 &  -0.020 \\
    4 &   3 \pm \hphantom{0}0.053 &   4.305 &   3 \pm \hphantom{0}0.059 &  2.975 &   3 \pm \hphantom{0}0.059 &   3.014 &   3 \pm \hphantom{0}0.059 &   3.017 \\
    5 &   0 \pm \hphantom{0}0.301 &  -3.550 &   0 \pm \hphantom{0}0.330 &  0.059 &   0 \pm \hphantom{0}0.327 &  -0.066 &   0 \pm \hphantom{0}0.329 &   0.020 \\
    6 &  15 \pm \hphantom{0}0.852 &  42.839 &  15 \pm \hphantom{0}0.939 & 14.405 &  15 \pm \hphantom{0}0.948 &  15.110 &  15 \pm \hphantom{0}0.941 &  15.052 \\
    7 &   0 \pm \hphantom{0}3.923 & -78.766 &   0 \pm \hphantom{0}4.324 &  0.658 &   0 \pm \hphantom{0}4.288 &   0.054  &   0 \pm \hphantom{0}4.300 &   3.077 \\
    8 & 105 \pm            14.070 & 671.864 & 105 \pm            15.591 & 98.687 & 105 \pm 15.727 & 107.839  & 105 \pm 15.577 & 106.745 \\
 \end{tabular*}
 \end{ruledtabular}
\end{table*}

A handy way of checking for the normality of the residuals is looking
into the standardized moments~\cite{evans2004probability}. These are
defined as
\begin{equation}
 \mu_r' = \frac{1}{N} \sum_{i=1}^N \left( \frac{X_i - \mu}{\sigma}\right)^r,
\end{equation}
where $\mu$ is the arithmetic mean and $\sigma$ the standard
deviation; the $r=1$ and $r=2$ standardized moments are zero and one
respectively. Due to the finite size of any random sample an intrinsic
uncertainty $\Delta \mu_r'(N)$ exist. This uncertainty can be
estimated using Monte Carlo simulations with $M$ random samples of
size $N$ and calculating the standard deviation of $\mu_r'$. The
result of such simulations are shown in table~\ref{tab:Cmoments} along
with the moments of the residuals of the complete database with
$N=8125$ data and the $3\sigma$ self-consistent database with
$N=6713$. Clearly, the complete database shows discrepancies at $68\%$
confidence level and hence cannot be attributed to the finite size of
the sample. On the other hand for the $3\sigma$ self-consistent
database the moments fall in the expected interval.  This is a first
indication on the validity of Eq.~(\ref{eq:generator}) for our fit to
this database.

The moments method was already used by the Nijmegen
group~\cite{Bergervoet:1988zz} for the available at the time pp (about
400) data up to $T_{\rm LAB}=30{\rm MeV}$. However, they tested the
squared residuals $R_i^2$ in Eq.~(\ref{eq:residuals}) with a
$\chi^2$-distribution with one degree of freedom which corresponds to
testing only even moments of the normal distribution.  As we have
already pointed out, this is insensitive to the sign of $R_i$ and
hence may overlook relevant skewness.

\subsection{Rescaling of errors}

The usefulness of the normality test goes beyond checking the
assumptions of the $\chi^2$ fit since it allows to extend the validity
of the method to naively unfavorable situations.

 Indeed, if the
actual value for $\chi^2_{\rm min}/\nu$ comes out outside the interval
$1\pm \sqrt{2/\nu}$ one can still re-scale the errors by the so-called
Birge factor~\cite{birge1932calculation} namely $\Delta O_i^{\rm exp}
\to \sqrt{\chi^2_{\rm min}/\nu} \Delta O_i^{\rm exp}$ so that the new
figure of merit is
\begin{eqnarray}
\bar \chi^ 2= (\chi^2/\chi^2_{\rm min}) \nu 
\end{eqnarray}
which by definition fulfills $\bar \chi^2_{\rm min}/\nu=1$.  There is
a common belief that this rescaling of $\chi^2$ restores normality,
when it only normalizes the resulting distribution~\footnote{This
  rescaling is a common practice when errors on the fitted quantities
  are not provided; uncertainties are invented with the condition that
  indeed $\chi^2_{\rm min}/\nu \sim 1$. The literature on phase-shift
  analyzes is plenty of such examples. It is also a recommended
  practice in the Particle Data Group booklet when incompatible data
  are detected among different sets of
  measurements~\cite{Beringer:1900zz,behnke2013data}).}.  If this was
the case, there is no point in rejecting any single datum from the
original database. Of course, it may turn out that one finds that
residuals are non-standardized normals. That means that they would
correspond to a scaled gauss distribution.  We will show that while
this rescaling procedure works {\it once} the residuals obey a
statistical distribution, the converse is not true; rescaling does not
make residuals obey a statistical distribution.

In the case at hand we find that rescaling only works for the
$3\sigma$-self consistent database because residuals turn out to be
normal. We stress that this is not the case for the full database.  Of
course, there remains the question on how much can errors be globally
changed by a Birge factor. Note that errors quoted by experimentalists
are in fact estimates and hence are subjected to their own
uncertainties which ideally should be reflected in the number of
figures provided in $\Delta O_i^{\rm exp}$. For $N \gg P $ one has
$\nu \sim N $ and one has $\chi^2/\nu = 1 \pm \sqrt{2/N}= 1\pm 0.016 $
for $N=8000$. Our fit to the complete database yields $\chi^2/\nu=1.4$
which is well beyond the confidence level.  Rescaling in this case
would correspond to globally enlarge the errors by $\sqrt{1.4} \sim
1.2$ which is a $20\%$ correction to the error in {\it all}
measurements. Note that while this may seem reasonable, the rescaled
residuals do not follow a Gaussian distribution. Thus, the noise on
Eq.~(\ref{eq:generator}) remains unknown and cannot be statistically
interpreted.

For instance, if we obtain $\chi^2/\nu=1.2$ one would globally enlarge
the errors by $\sqrt{1.1} \sim 1.1$ which is a mere $10\%$ correction
on the error estimate, a perfectly tolerable modification which
corresponds to quoting just one significant figure on the
error~\footnote{For instance, quoting $12.23(4) \equiv 12.23\pm0.04$
  means that the error could be between 0.035 and 0.044 which is
  almost $25\%$ uncertainty in the error. Quoting instead 12.230(12)
  corresponds to a $10\%$ uncertainty in the error.}. Thus, while
$\chi^2_{\rm min}/\nu=1\pm \sqrt{2/\nu}$ looks as a sufficient
condition for goodness of fit, it actually comes from the assumption
of normality of residuals. However, one should not overlook the
possibility that the need for rescaling might in fact suggest the
presence of unforeseen systematic errors.

\begin{table}
 \caption{\label{tab:PearsontestResults} Results of the Pearson 
 normality test of the residuals obtained by fitting the complete data
 base with a delta-shell potential and the $3\sigma$ consistent database
 with the delta-shell and the OPE-Gaussian potentials. The results of
 the test of the scaled residuals for every case is shown below the
 corresponding line. The critical value $T_c$ corresponds to a
 significance level of $\alpha=0.05$.}
 \begin{ruledtabular}
 \begin{tabular*}{\columnwidth}{@{\extracolsep{\fill}} c c c D{.}{.}{2.3} D{.}{.}{3.2} D{.}{.}{1.9}}
   Database  & Potential  &$N$ &  \multicolumn{1}{c}{$T_{c}$} & \multicolumn{1}{c}{$T_{\rm obs}$} & \multicolumn{1}{c}{$p$-value}  \\
  \hline  
   Complete  & OPE-DS      & $8125$ & 93.945 & 598.84 & 1.36\times10^{-83}   \\
               &    &        &        & 190.16 & 2.18\times10^{-12}   \\
   $3\sigma$ & OPE-DS    & $6713$ & 87.108 &  82.67 & 0.09   \\
               &    &        &        &  69.08 & 0.40    \\
   $3\sigma$ & $\chi$TPE-DS    & $6712$ & 87.108 &  100.70 & 0.004   \\
               &    &        &        &  74.40 & 0.25    \\
   $3\sigma$ & OPE-G & $6711$ & 87.108 &  84.17 & 0.08   \\
               &    &        &        &  68.38 & 0.43    \\
 \end{tabular*}
 \end{ruledtabular}
\end{table}

\section{Normality tests for residuals}
\label{sec:normal}

There is a large body of statistical tests to quantitatively assess
deviations from an specific probability distribution (see
e.g. \cite{conover1980practical}). In these procedures the
distribution of empirical data $X_i$ is compared with a theoretical
distribution $F_0$ to test the \emph{null hypothesis} $H_0: X_i \sim
F_0$. If statistically significant differences are found between the
empirical and theoretical distributions the null hypothesis is
rejected and its negation, the \emph{alternative hypothesis} $H_1: X_i
\sim F_1$ is considered valid, where $F_1$ is an unknown distribution
different from $F_0$. The comparison is made by a \emph{test
  statistic} $T$ whose probability distribution is known when
calculated for random samples of $F_0$; different methods use
different test statistics. A decision rule to reject (or fail to
reject) $H_0$ is made based on possible values of $T$, for example if
the observed value of the test statistic $T_{\rm obs}$ is greater (or
smaller depending on the distribution of $T$) than a certain critical
value $T_c$ the null hypothesis is rejected. $T_c$ is determined by
the probability distribution of $T$ and the desired \emph{significance
  level} $\alpha$, which is the maximum probability of rejecting a
true null hypothesis. Typical values of $\alpha$ are $0.05$ and
$0.01$. Another relevant and meaningful quantity in hypothesis testing
is the $p$-value, which is defined as the smallest significance level
at which the null hypothesis would be rejected. Therefore a small
$p$-value indicates clear discrepancies between the empirical
distribution and $F_0$. A large $p-$value, on the contrary, means that
the test could not find significant discrepancies.

In our particular case $H_0$ is that the residuals follow a standard
normal distribution, and the $p$-value would be the probability that
denying the assumption of true normality would be a erroneous
decision.

\subsection{Pearson-Test}

A simple way of testing the goodness of fit is by using the Pearson
test by computing the test statistic
\begin{eqnarray}
T=\sum_{i=1}^{N_b} \frac{(n_i^{\rm fit}-n_i^{\rm normal})^2}{n_i^{\rm th}}
\end{eqnarray}
where $n_i^{\rm fit}$ are the number of residuals on each bin and
$n_i^{\rm normal}$ are the number of expected residuals for the normal
distribution in the same bin. $T$ follows a $\chi^2$-distribution with
$N_b-1$ d.o.f. The decision on how close is a given histogram the
expected distribution depends on the specific choice of binning, which
is the standard objection to this test. To perform the test we use a
equiprobable binning so that $\Delta R_i$ is such that $n_i^{\rm
  normal}$ is constant for all $i$, instead of the equidistant binning
shown in Fig.~\ref{Fig:Histograms} (see
e.g. Ref.~\cite{eadie2006statistical} for more details on binning
strategies). The results of the test are given in
Table~\ref{tab:PearsontestResults} and, as we see, {\it again} the
complete data base fails the test even when residuals are scaled.

\subsection{Kolmogorov-Smirnov-Test}

A simple and commonly used test is the Kolmogorov-Smirnov (KS)
test~\cite{kolmogorov1933sulla, smirnov1948}. The KS test uses the empirical
distribution function $S(x)$ defined as the fraction of $X_i$s that are
less or equal to $x$ and expressed by
\begin{equation}
 S(x) = \frac{1}{N}\sum_{i=1}^N \theta(x-X_i),
\end{equation}
where $N$ is the number of empirical data. The test statistic in this 
procedure is defined as the greatest difference between $S(x)$ and $F_0(x)$,
that is
\begin{equation}
 T_{\rm KS} = \sup_{x}|F_0(x) - S(x)|.
\end{equation}
Some of the advantages of using $T_{\rm KS}$ as a test statistic come
from its distribution under the null hypothesis, since it is
independent of $F_0$, it can be calculated analytically and a fairly
good approximation exists for the case of large $N$. Given that large
values of $T_{\rm KS}$ indicate large deviations from the theoretical
distribution the decision rule will be to reject the null hypothesis
if the observed value $T_{\rm obs, KS}$ is larger than a certain
critical value $T_{c, \rm KS}$. The critical value depends on $\alpha$
and $N$; for large number of data and a significance level of $0.05$
$T_{c,\rm KS} = 1.36/\sqrt{N}$. Also, a good approximation for the
corresponding $p$-value has been given~\cite{von1964mathematical}
\begin{equation}
P_{\rm KS}(T_{\rm obs}) = 2 \sum_{j=1}^{\infty} (-1)^{j-1} e^{-2
\left[\left(\sqrt{N}+0.12+0.11/\sqrt{N}\right)jT_{\rm obs}\right]^2}.
\end{equation}

\begin{table}
 \caption{\label{tab:KStestResults} Same as
 table~\ref{tab:PearsontestResults} for the Kolmogorov-Smirnov test}
 \begin{ruledtabular}
 \begin{tabular*}{\columnwidth}{@{\extracolsep{\fill}} c c c D{.}{.}{1.3} D{.}{.}{1.3} D{.}{.}{1.9}}
   Database & Potential  & $N$ &  \multicolumn{1}{c}{$T_{c}$} & \multicolumn{1}{c}{$T_{\rm obs}$} & \multicolumn{1}{c}{$p$-value}  \\
  \hline  
   Complete  & OPE-DS      & $8125$ & 0.015 & 0.037 & 4.93\times10^{-10} \\
             &      &        &       & 0.035 & 6.24\times10^{-9}  \\
   $3\sigma$ & OPE-DS    & $6713$ & 0.017 & 0.011 & 0.43   \\
             &      &        &       & 0.012 & 0.26   \\
   $3\sigma$ & $\chi$TPE-DS    & $6712$ & 0.017 & 0.010 & 0.47   \\
             &      &        &       & 0.010 & 0.47   \\
   $3\sigma$ & OPE-G & $6711$ & 0.017 & 0.013 & 0.22   \\
             &      &        &       & 0.014 & 0.18    \\
 \end{tabular*}
 \end{ruledtabular}
\end{table}

\begin{figure}[htbp]
\begin{center}
\epsfig{figure=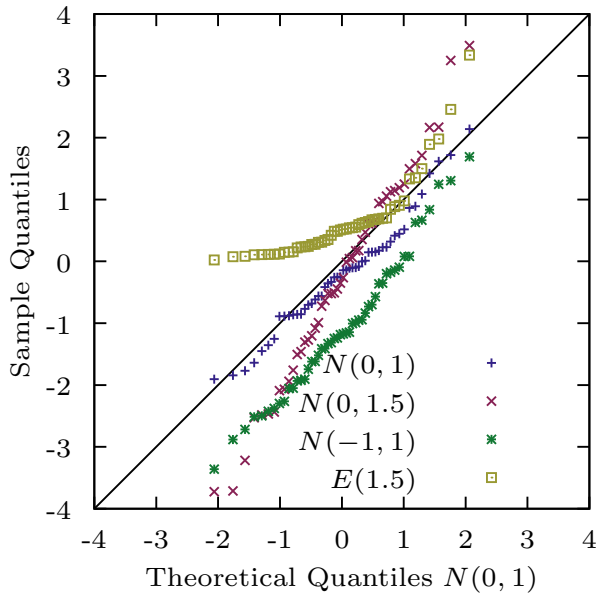,width=0.9\linewidth}  
\end{center}
\caption{(Color online) Quantile-Quantile plot of different random samples
against the standard normal distribution. Blue crosses are sampled from 
the $N(0,1)$ distribution, red diagonal crosses from $N(0,1.5)$, green
asterisks from $N(-1,1)$ and yellow squares from the exponential distribution
$E(1.5)$}
\label{Fig:QQplotExamples}
\end{figure}

The results of the KS normality test to the residuals obtained by
fitting the potential parameters to the complete and $3\sigma$
consistent databases are shown in table~\ref{tab:KStestResults}.
For the case of the complete database the observed test statistic is
much larger than the critical value at the $0.05$ significance level
which indicates that with a $95 \%$ confidence level the null hypothesis
$H_0: X_i \sim N(0,1)$ can be rejected; the extremely low $p$-value 
gives an even greater confidence level to the rejection of $H_0$ very
close to the $100 \%$. In contrast the observed test statistic using the
$3\sigma$ consistent data is smaller than the corresponding critical
value, this indicates that there is no statistically significant 
evidence to reject $H_0$.

A shortcoming of the KS test is that the sensitivity to deviations from
$F_0(x)$ is not independent from $x$. In fact the KS test is most
sensitive to deviations around the median value of $F_0$ and therefore
is a good test for detecting shifts on the probability distribution,
which in practice are unlikely to occur in the residuals of a least
squares fit. But in turn, discrepancies away from the median such as
spreads, compressions or outliers on the tails, which are not that
uncommon on residuals, may go unnoticed by the KS test.

\subsection{Quantile-Quantile plot}

\begin{figure}[htbp]
\begin{center}
\epsfig{figure=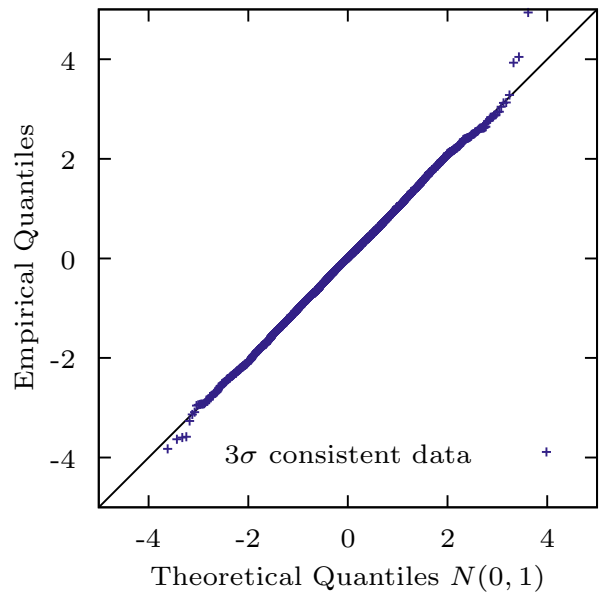,width=0.9\linewidth}  
\end{center}
\caption{(Color online) Quantile-Quantile plot of the residuals
  obtained from fitting the $3\sigma$ consistent database against the
  standard normal distribution. The deviations at the tails, which are
  not detected using the Kolmogorov-Smirnov test, are clearly visible
  with this graphical tool.}
\label{Fig:QQresiduals}
\end{figure}

A graphical tool to easily detect the previously mentioned
discrepancies is the quantile-quantile (QQ) plot which maps two
distributions quantiles against each other. The $q$-quantiles of a
probability distribution are obtained by taking $q-1$ equidistant
points on the $(0,1)$ interval and finding the values whose cumulative
distribution function correspond to each point. For example, to find
the $4$-quantiles of the normal distribution with zero mean and unit
variance we take the points $0.25$, $0.5$ and $0.75$ and look for
values of $x$ satisfying
\begin{equation}
\frac{1}{\sqrt{2\pi}} \int_{-\infty}^{x}{e^{-\frac{-\tilde{x}}{2}}d
\tilde{x}} = 0.25, \ 0.5, \ 0.75.
\end{equation}
In this case the $4$-quantiles are $-0.6745$, $0$ and $0.6745$. For a
set of ranked empirical data the easiest way to find the $q$-quantiles
is to divide it into $q$ essentially equal sized subsets and take the
$q-1$ boundaries as the quantiles.

To compare empirical data with a theoretical distribution function using
a QQ plot the $N+1$-quantiles are used. In this way each data can be
graphed against the corresponding theoretical distribution's quantile;
if the empirical and theoretical distributions are similar the QQ plot
points should lie close to the $y=x$ line. In
Fig.~\ref{Fig:QQplotExamples} different random samples of size $N=50$
are compared with a normal distribution. The first sample corresponds to
the $N(0,1)$ distribution, the second to the $N(0,1.5)$ and the larger
spread of the data can be seen as a shift on the tails towards the
bottom left and top right parts of the graph. A third samples comes from
the $N(-1,1)$ distribution and this can be seen as an downward shift of
the points. A last sample is taken from the exponential distribution
$E(1.5)$ which is asymmetric and positive.

Fig.~\ref{Fig:QQresiduals} shows the QQ plot of the residuals from the
fit to the $3 \sigma$ consistent database against the $N(0,1)$
distribution; deviations around the tails, which can not be seen with
the histogram in Fig.~\ref{Fig:Histograms} and are not detected by the
Pearson and KS tests, are clearly visible at the bottom left and
top right corners of the plot.

\begin{figure*}[htbp]
\begin{center}
\epsfig{figure= 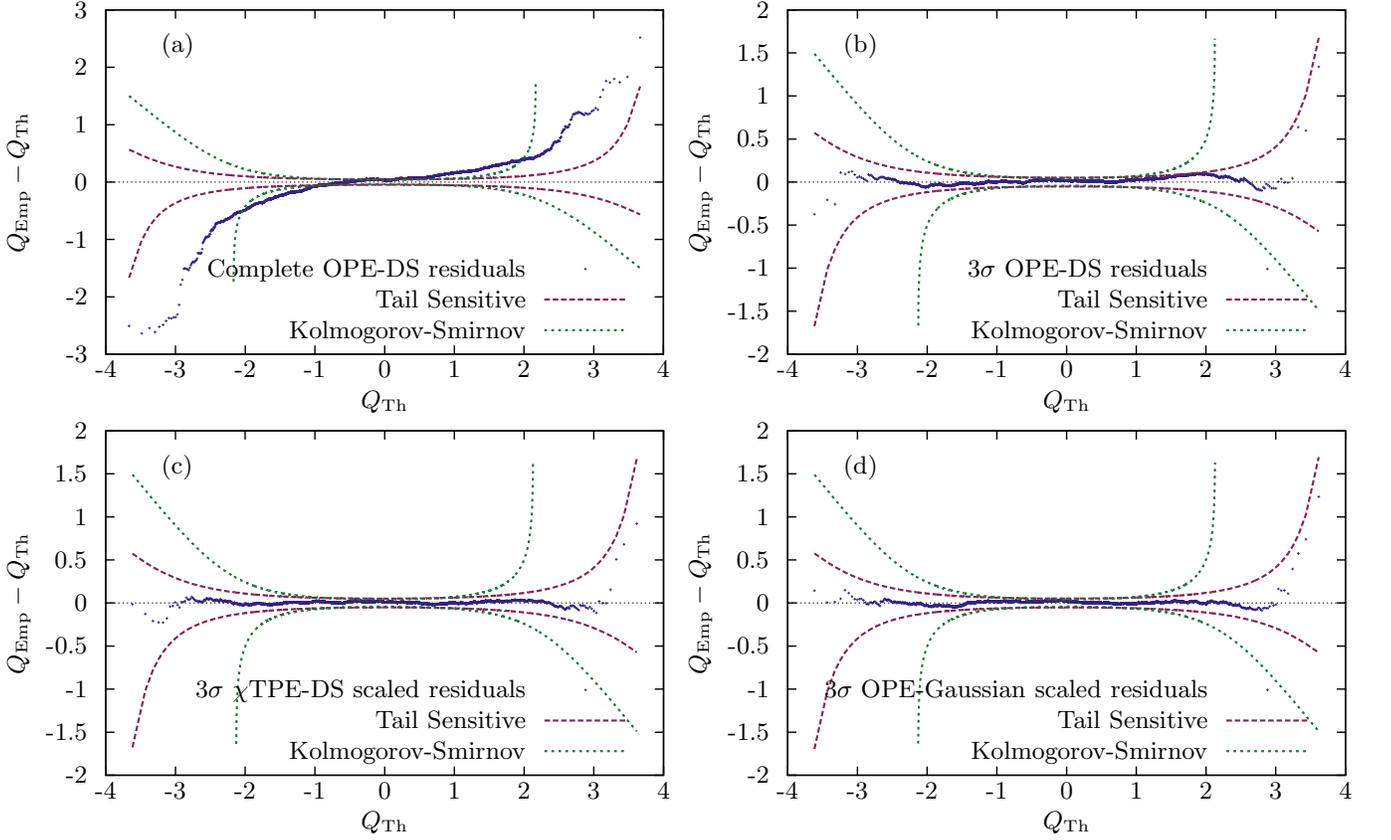,width=\linewidth}  
\end{center}
\caption{(Color online) Rotated Quantile-Quantile plot of the
  residuals obtained (blue points) from fitting the complete database
  with the OPE-Delta-Shell potential (upper left panel), the $3\sigma$
  self-consistent database fitted with the OPE-Delta-Shell potential
  (upper right panel), the $\chi$TPE-Delta-Shell potential (lower left
  panel) and the OPE-Gaussian potential (lower right panel). $95\%$
  confidence bands of the TS (red dashed lines) and KS (green dotted
  lines) tests are included.}
\label{Fig:QQconfidence}
\end{figure*}

\subsection{Tail-Sensitive-Test}

\begin{table}
 \caption{\label{tab:TStestResults} Same as
 table~\ref{tab:PearsontestResults} for the tail sensitive test}
 \begin{ruledtabular}
 \begin{tabular*}{\columnwidth}{@{\extracolsep{\fill}} c c c D{.}{.}{1.5} D{.}{.}{1.9} D{.}{.}{3.4}}
   Database & Potential & $N$ &  \multicolumn{1}{c}{$T_{c}$} & \multicolumn{1}{c}{$T_{\rm obs}$} & \multicolumn{1}{c}{$p$-value}  \\
  \hline  
   Complete & OPE-DS       & $8125$ & 0.00070 & 0.0000              & < 0.0002 \\
            &       &        &         & 3.54\times10^{-25}  & < 0.0002  \\
   $3\sigma$ & OPE-DS    & $6713$ & 0.00072 & 0.0010              & 0.07   \\
             &      &        &         & 0.0076              & 0.32   \\
   $3\sigma$ & $\chi$TPE-DS    & $6712$ & 0.00072 & 0.0005              & 0.03   \\
             &      &        &         & 0.0156              & 0.50   \\
   $3\sigma$ & OPE-G & $6711$ & 0.00072 & 0.0001             & 0.01   \\
             &      &        &         & 0.0082              & 0.33    \\
 \end{tabular*}
 \end{ruledtabular}
\end{table}

Even though the QQ plot is a convenient and easy-to-use tool to detect
deviations from a theoretical distribution, graphical methods often
depend on subjective impressions and no quantitative description of
the deviations visible in Fig.~\ref{Fig:QQresiduals} can given by the
QQ plot alone. A recent method by Aldor-Noiman \emph{et
  al.}~\cite{Aldor2013} provides $(1-\alpha)$ confidence bands to the
QQ plot to quantitatively test deviations from the normal
distribution. This new test, called tail sensitive (TS), has a higher
sensitivity on the tails than the KS test. In fact, the TS test
rejection rate is uniformly distributed over the $x$
variable. Although no analytic expression is given for the TS test
statistic distribution, it can be easily simulated via Monte-Carlo
techniques. The details of such simulation are explained
in~\cite{Aldor2013}.  We will restrict ourselves to point out that, a
small value of $T_{\rm TS}$ indicates discrepancies between the
empirical and normal distribution and therefore the rejection
criterion for the null hypothesis is $T_{\rm obs, TS} < T_{c, \rm
  TS}$~\footnote{It should also be noted that a typo in
  Ref.~\cite{Aldor2013} is made in their steps 1c and 1e where
  $\Phi^{-1}$ and $B_{(i,n+1-i)}^{-1}$ are printed instead of $\Phi$
  and $B_{(i,n+1-i)}$; the latter are consistent with the rest of the
  text and the results presented there.}

We applied the TS normality test to both sets of residuals, the
complete database and the $3\sigma$ consistent one, and show the
results on table~\ref{tab:TStestResults}. For each test the
Monte-Carlo simulation consisted on taking $5000$ random samples of
size $N$ with a standard normal distribution and calculating $T_{\rm
  obs, TS}^{\rm MC}$ for each sample to obtain the distribution of
$T_{\rm TS}$ under the null hypothesis. The critical value for a
significance level $\alpha = 0.05$ corresponds to the $T_{\rm obs,
  TS}^{\rm MC}$ that is greater than $5\%$ of all the values
calculated. Finally the test statistic for the empirical data $T_{\rm
  obs, TS}^{\rm emp}$ can be calculated and compared to the simulated
distribution to obtain the $p$-value. In this case the $p$-value is
the proportion of $T_{\rm obs, TS}^{\rm MC}$ that are smaller than
$T_{\rm obs, TS}^{\rm emp}$. Since the observed $T_{TS}$ for the
complete database residuals is numerically equal to zero and smaller
than all of the simulated values we can only give an upper bound to
the $p$-value.  The graphical results of the TS test are presented in
Fig.~\ref{Fig:QQconfidence} with the $95\%$ confidence level bands;
the same bands for the KS test are drawn for comparison reasons. Since
for such a large value of $N$ the confidence bands are very narrow, a
$45^o$-clockwise rotated QQ plot is used to visually enhance the
possible deviations from a normal distribution. The complete database
residuals (upper left panel) show obvious deviations from the normal
distribution which is reflected on the extremely low $p$-values. The
$3\sigma$ consistent data residuals (upper right panel) show
deviations from the normal distribution that are always within the TS
confidence bands and therefore to a confidence level $\alpha=0.05$
there are no statistically significant differences to reject the null
hypothesis.

\begin{figure}
\begin{center}
\epsfig{figure=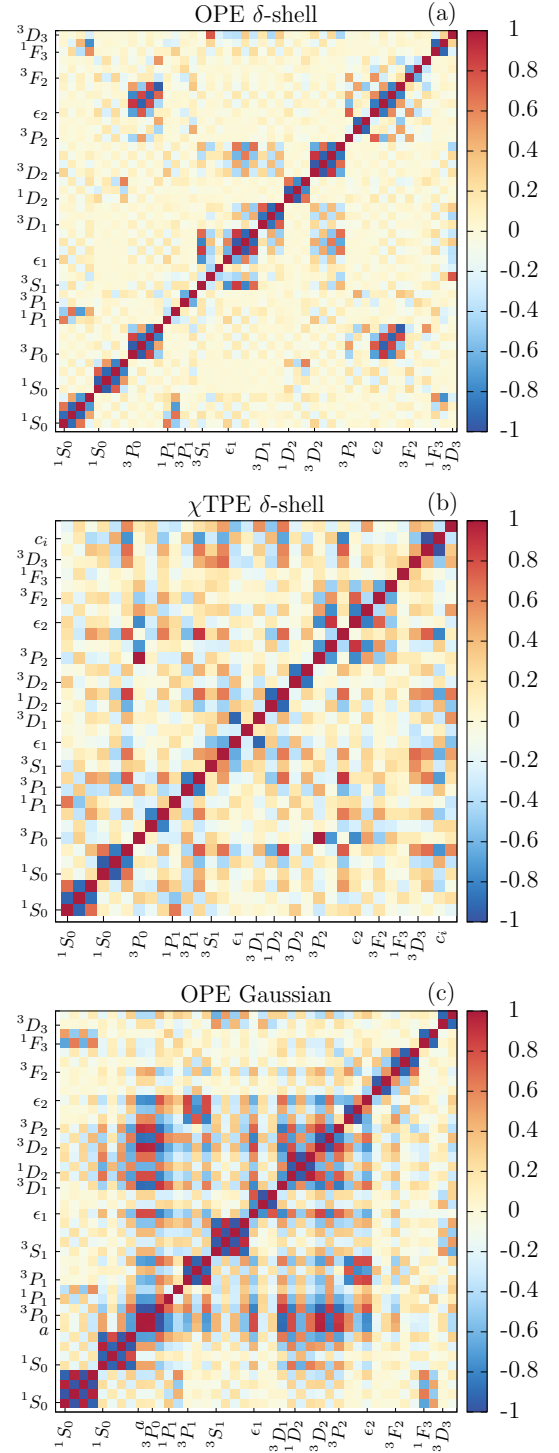,width=.8\columnwidth}
\end{center}
\caption{(Color online) Correlation matrix ${\cal C}_{ij}$ for the
  short distance parameters in the partial wave basis
  $(V_i)^{LSJ}_{l,l'}$, see Eq.~(\ref{eq:potential-short}). We show
  the OPE-DS (upper panel) and the $\chi$TPE-DS (middle panel)
  potentials. The points $r_i = \Delta r_\pi (i+1)$ are grouped within
  every partial wave. The ordering of parameters is as in the
  parameter tables in Refs.~\cite{Perez:2013mwa,Perez:2013jpa}
  and~\cite{Perez:2013oba} for OPE-DS 46 parameters and the $\chi$TPE
  30+3 parameters (the last three are the chiral constants
  $c_1,c_3,c_4$) respectively. The OPE-Gaussian case (lower panel)
  also contains the parameter $a$. We grade gradually from $100\%$
  correlation, ${\cal C}_{ij}=1$ (red), $0\%$ correlation, ${\cal
    C}_{ij}=0$ (yellow) and $100\%$ anti-correlation, ${\cal
    C}_{ij}=-1$ (blue).}
\label{fig:corr}
\end{figure}

\subsection{Discussion}

We have shown in the previous discussion evidence supporting the
validity of Eq.~(\ref{eq:generator}) for the $3\sigma$-self consistent
database recently built from all published np and pp scattering data
since 1950 till 2013~\cite{NavarroPerez:2011fm,Perez:2013cza}. The
numerics can be a costly procedure since multiple optimizations must
be carried out and different subsets of data of the complete database
must be tested and confronted. As outlined above, our analysis was
carried out using a physically motivated coarse grained potential and
more specifically a delta-shells interaction already proposed by
Avil\'es~\cite{Aviles:1973ee}.  This scheme proved extremely
convenient for fast minimization and error evaluation.  

As a first application, with the currently fixed database we have also
addressed the calculation of the chiral constants which appear in the
$\chi$TPE potential~\cite{Perez:2013oba} which also passes the
normality test as can be seen from Fig.~\ref{Fig:QQconfidence} and
Tables~\ref{tab:Cmoments}, \ref{tab:PearsontestResults},
\ref{tab:KStestResults} and \ref{tab:TStestResults}).  We note that
the small rescaling by the Birge factor $\sqrt{1.07}$ is requested to
pass the Pearson and TS tests. As we have mentioned, this form of
$\delta$-shell potentials cannot be directly implemented in some of
the many powerful computational approaches to nuclear structure
calculations~\footnote{The $\delta$-shell potential cannot even be
  plotted, which may naively seem a disadvantage. However, its Fourier
  transformation is smooth~\cite{NavarroPerez:2011fm} in the relevant
  CM momentum region of $p_{\rm CM} \lesssim 2 {\rm fm}^{-1}$,
  complying to the idea that coarse graining down to $\Delta r_\pi
  \sim 0.6 {\rm fm}$ resolutions lacks information on shorter length
  scales.}.

\begin{figure*}
\begin{center}
\epsfig{figure=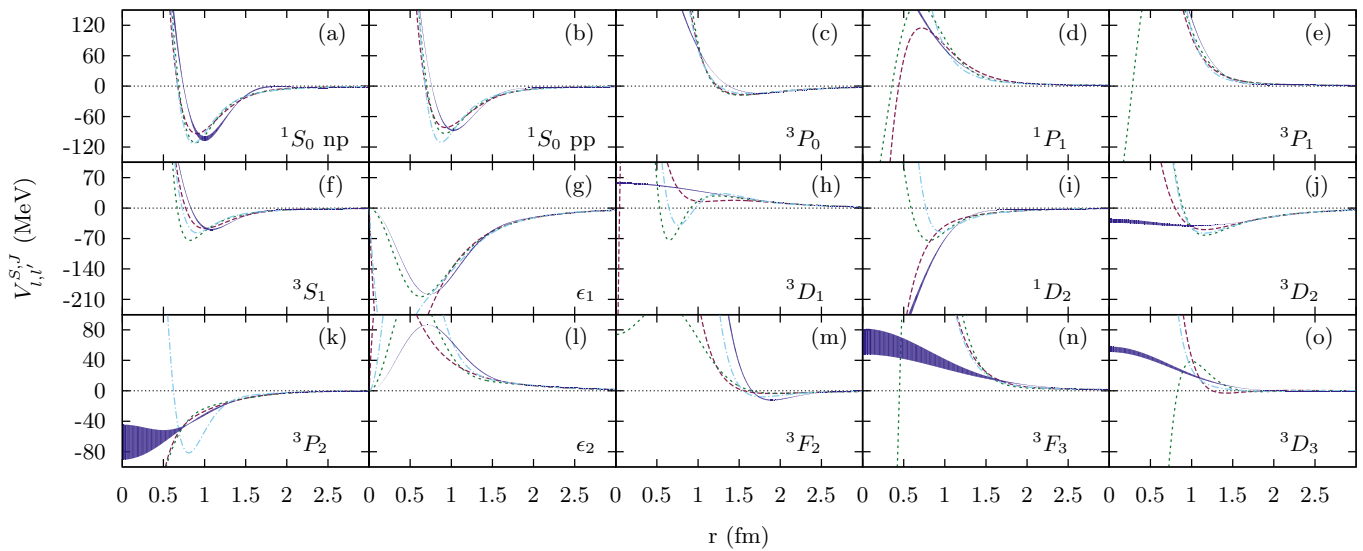,width=\linewidth}
\end{center}
\caption{(Color online) Lowest np and pp partial waves potentials (in
  MeV) and their errors (solid band) as a function of the
  inter-nucleon separation (in fm) for the present OPE+Gaussian
  analysis (blue band) Reid93~\cite{Stoks:1994wp} (red dashed)
  NijmII~\cite{Stoks:1994wp} (green dotted) and
  AV18~\cite{Wiringa:1994wb} (light-blue dashed-dotted) as a function
  of the inter-nucleon distance $r$ (in fm).}
\label{fig:local-pot}
\end{figure*}

\begin{figure*}
\begin{center}
\epsfig{figure=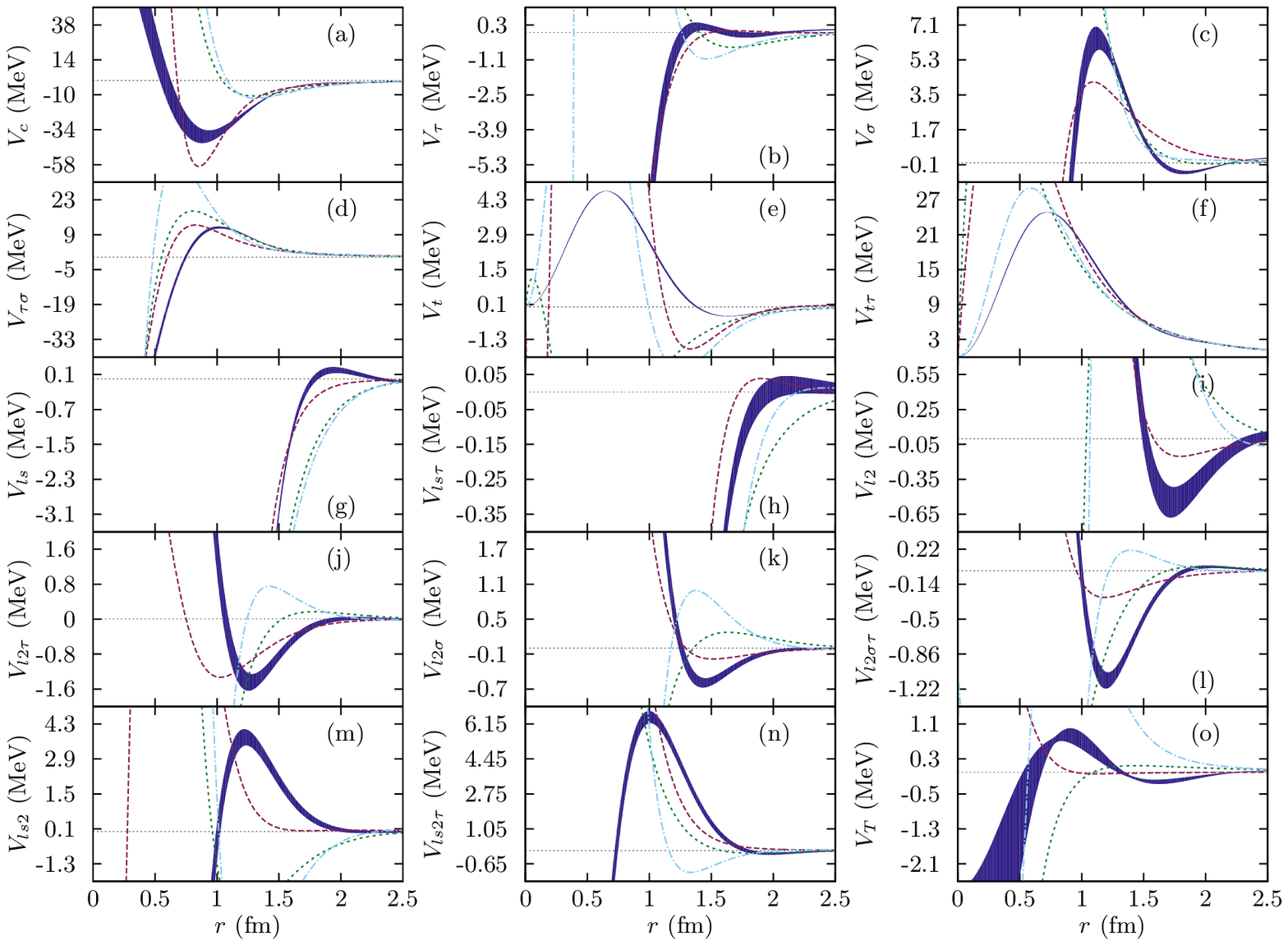,width=\linewidth}
\end{center}
\caption{(Color online) NN potentials (in MeV) in the operator basis
  with errors (solid band) as a function of the inter-nucleon
  separation (in fm) for the present OPE+Gaussian analysis (blue band)
  Reid93~\cite{Stoks:1994wp} (red dashed) NijmII~\cite{Stoks:1994wp}
  (green dotted) and AV18~\cite{Wiringa:1994wb} (light-blue
  dashed-dotted) as a function of the inter-nucleon distance $r$ (in
  fm).}
\label{fig:operator-pot}
\end{figure*}

The necessary conditions for a sensible interpretation of the $\chi^2$
fit according to Eq.~(\ref{eq:generator}) requires testing for
normality of residuals of a fit to a consistent database.  In all, the
present situation regarding both the selection of data with the
self-consistent $3\sigma$ criterion and the normality of residuals
turns out to be highly satisfactory. In our view, this combined
consistency of the statistical assumptions and the theory used to
analyze it provides a good starting point to proceed further in the
design of theory-friendly smooth NN interactions as well as a sound
estimate of their statistical uncertainties.

Of course, the normality of residuals applies to {\it any} fit aiming
at representing the data. Thus, any potential which pretends to
represent the data ought to pass the test. In the next section we
propose a potential whose short distance part is made of a
superposition of Gaussian functions and, unlike the $\delta$-shell
potential, can be plotted.  We will check that our proposed potential
does in fact pass the normality test.

There is an issue concerning the statistical approach on {\it what}
would be the ``true'' potential since the concept of true parameters
of a given model is invoked (see the discussion in
Sect~\ref{sec:modeling}). On the one hand, the very definition of
potential is subject to ambiguities because the scattering information
only determines an interaction once its specific form has been
chosen~\cite{chadan2011inverse}. This reflects the well known
off-shell ambiguities which by definition are inaccessible to
experiment~\cite{srivastava1975off}. On the other hand, nuclear
structure calculations are carried out with potentials statistically
representing the scattering data. This is a source for a systematic
uncertainty which was unveiled in
Ref.~\cite{NavarroPerez:2012vr,Perez:2012kt,Amaro:2013zka} for the
previously developed high-quality interactions. The upgrade of this
systematic uncertainty study using the present statistical analysis is
left for future research.

Ultimately, QCD is the theory to validate Eq.~(\ref{eq:generator}) vs
the large body of data, $O_i^{\rm th}=O_i^{\rm QCD}$ with just two
parameters in the $(u,d)$ sector, $\Lambda_{\rm QCD}$ and the quark
masses $(m_u,m_d)$, or equivalently with the pion weak decay constant
$f_\pi$ and the pion masses $(m_{\pi^0},m_{\pi^\pm})$. Remarkably,
nuclear potentials have been evaluated on the lattice
recently~\cite{Aoki:2009ji,Aoki:2011ep,Aoki:2013tba}. The HAL QCD
Collaboration~\cite{HALQCD:2012aa} finds a local potential for the
unphysical pion mass $m_\pi=701 {\rm MeV}$ with a shape similar to our
OPE-Gaussian potential (see Sect.~\ref{sec:OPEgauss}) but a depth of
$-30 {\rm MeV}$ in the central component $V_c$ and $\Delta V_c \sim 5
{\rm MeV}$ for $r \gtrsim 1 {\rm fm}$, and consequently the $^ 1S_0$
phase-shift obtained by directly solving the Schr\"odinger equation is
smaller as compared to ours with much larger errors. This potential
approach uses the Nambu-Bethe-Salpeter wave function which ultimately
depends on the choice of the interpolating composite nucleon fields
(for a recent overview of the pros and cons of the potential approach
to lattice QCD see e.g.~\cite{Walker-Loud:2014iea}). Of course, since
the lattice NN potential depends ultimately in just two parameters,
$\Lambda_{\rm QCD}$ and $m_q$ the different r-values in the potential
functions $V_n (r)$ {\it must be} correlated. In the phenomenological
approach correlations among the fitting parameters are indeed found or
built in. Some of them are the trivial ones due to the OPE-potential
which just depends on the pion masses $(m_{\pi^0},m_{\pi^\pm})$, but
others correspond to the inner short distance parameters, suggesting
that the number of parameters can de reduced {\it solely} from the
phenomenological potential analysis of the data. In
Fig.~\ref{fig:corr} we represent pictorially the resulting correlation
matrix both for the OPE-DS fit~\cite{Perez:2013mwa,Perez:2013jpa} as
well as for $\chi$TPE-DS~\cite{Perez:2013oba} short distance
parameters in the partial wave basis $(V_i)^{LSJ}_{l,l'}$, see
Eq.~(\ref{eq:potential-short}). The listing ordering is the same as
the one in the parameter tables in
Refs.~\cite{Perez:2013mwa,Perez:2013jpa} and~\cite{Perez:2013oba} for
OPE-DS and $\chi$TPE-DS respectively.  Note, the isolated pattern of
correlations for the OPE-DS case, however as we see there are
substantial correlations among different $(V_i)^{LSJ}_{l,l'}$ within a
given partial wave suggesting the possibility of reducing the number
of parameters.  Indeed, we observe that this parameter reduction takes
place from $46$ to $33$ when going from the OPE-DS case to the
$\chi$TPE-DS potential~\cite{Perez:2013oba} which incorporates
specific QCD features such as chiral symmetry. The resulting
correlation pattern becomes now more spread over the full short
distance parameter space.

\begin{table}
 \caption{\label{tab:FitParameters} Fitting partial wave parameters
   $(V_i)^{JS}_{l,l'} $ (in MeV) with their errors for all states in
   the $JS$ channel. $-$ indicates that the corresponding fitting
   $(V_i)^{JS}_{l,l'} =0$. The parameters marked with $^*$ are set to
   have the tensor components vanish at the origin. The parameter
   $a$, which determines the width of each Gaussian, is also used as a
   fitting parameter and the value $2.3035 \pm 0.0133$ fm is found.}
 \begin{ruledtabular}  
 \begin{tabular*}{\textwidth}{@{\extracolsep{\fill}}l D{.}{.}{4.4} D{.}{.}{4.4} D{.}{.}{5.4} D{.}{.}{4.5} }
  Wave  & \multicolumn{1}{c}{$V_1$} & 
          \multicolumn{1}{c}{$V_2$} & 
          \multicolumn{1}{c}{$V_3$} & 
          \multicolumn{1}{c}{$V_4$}   \\
 \hline\noalign{\smallskip}
  $^1S_0{\rm np}$&    -67.3773 &      598.4930 &    -2844.7118 &     3364.9823 \\
                 & \pm  4.8885 & \pm   64.8759 & \pm  245.3275 & \pm  268.9192 \\
  $^1S_0{\rm pp}$&    -52.0676 &      408.7926 &    -2263.1470 &     2891.2494 \\
                 & \pm  1.1057 & \pm   12.9206 & \pm   57.0254 & \pm   76.3709 \\
  $^3P_0$        &    -60.3589 &\multicolumn{1}{c}{$-$}&      520.5645 &\multicolumn{1}{c}{$-$}\\
                 & \pm  1.2182 &               & \pm   17.4210 &               \\
  $^1P_1$        &     22.8758 &\multicolumn{1}{c}{$-$}&      256.2909 &\multicolumn{1}{c}{$-$}\\
                 & \pm  0.9182 &               & \pm    8.1078 &               \\
  $^3P_1$        &     35.6383 &     -229.1500 &      928.1717 &\multicolumn{1}{c}{$-$}\\
                 & \pm  0.9194 & \pm    9.0104 & \pm   28.8275 &               \\
  $^3S_1$        &    -42.4005 &      273.1651 &    -1487.4693 &     2064.7996 \\
                 & \pm  2.1344 & \pm   24.1462 & \pm   91.3195 & \pm  105.4383 \\
  $\varepsilon_1$&   -121.8301 &      262.7957 &    -1359.3473 &     1218.3817^* \\
                 & \pm  3.2650 & \pm   19.0432 & \pm   50.9369 & \pm   34.8398 \\
  $^3D_1$        &     56.6746 &\multicolumn{1}{c}{$-$}&\multicolumn{1}{c}{$-$}&\multicolumn{1}{c}{$-$}\\
                 & \pm  1.3187 &               &               &               \\
  $^1D_2$        &    -44.4366 &      220.5642 &     -617.6914 &\multicolumn{1}{c}{$-$}\\
                 & \pm  1.2064 & \pm   10.8326 & \pm   27.1533 &               \\
  $^3D_2$        &   -107.3859 &       74.8901 &\multicolumn{1}{c}{$-$}&\multicolumn{1}{c}{$-$}\\
                 & \pm  2.9384 & \pm    7.1627 &               &               \\
  $^3P_2$        &    -10.4319 &\multicolumn{1}{c}{$-$}&     -170.3098 &      132.4249 \\
                 & \pm  0.3052 &               & \pm    7.3280 & \pm   13.2310 \\
  $\varepsilon_2$&     50.0324 &     -177.7386 &      748.5717 &     -620.8659^* \\
                 & \pm  0.8985 & \pm    8.2027 & \pm   34.7849 & \pm   27.2518 \\
  $^3F_2$        &      6.3917 &     -659.4308 &     3903.1138 &\multicolumn{1}{c}{$-$}\\
                 & \pm  2.6615 & \pm   41.3707 & \pm  187.9877 &               \\
  $^1F_3$        &     28.5198 &       42.9715 &\multicolumn{1}{c}{$-$}&\multicolumn{1}{c}{$-$}\\
                 & \pm  3.0801 & \pm   19.5127 &               &               \\
  $^3D_3$        &     -9.6022 &       65.9632 &\multicolumn{1}{c}{$-$}&\multicolumn{1}{c}{$-$}\\
                 & \pm  0.8870 & \pm    4.3677 &               &               \\
 \end{tabular*}
 \end{ruledtabular}
\end{table}
\begin{table}[htb]
 \caption{\label{tab:OperatorParameters} Operator coefficients
   $V_{i,n}$ (in MeV) with their errors for the OPE-Gaussian
   potential.  The operators $t T$, $\tau z$ and $\sigma \tau z$ are
   set to zero}
 \begin{ruledtabular}  
 \begin{tabular*}{\textwidth}{@{\extracolsep{\fill}}l D{.}{.}{3.4} D{.}{.}{3.4} D{.}{.}{4.4} D{.}{.}{4.4} }
  Operator  & \multicolumn{1}{c}{$V_1$} & 
          \multicolumn{1}{c}{$V_2$} & 
          \multicolumn{1}{c}{$V_3$} & 
          \multicolumn{1}{c}{$V_4$}   \\
 \hline\noalign{\smallskip}
$c$             &    -19.2829 &    126.2986 &   -648.6244 &    694.4340 \\
                & \pm  0.6723 & \pm  7.7913 & \pm 33.1067 & \pm 36.8638 \\
$\tau$          &      2.3602 &    -25.4755 &    130.0301 &   -284.7219 \\
                & \pm  0.4287 & \pm  5.4291 & \pm 20.0608 & \pm 19.8417 \\
$\sigma$        &      6.0528 &    -75.1908 &    372.4133 &   -530.8121 \\
                & \pm  0.4311 & \pm  5.2742 & \pm 19.5580 & \pm 22.4309 \\
$\tau \sigma$   &      7.3632 &    -48.5435 &    273.7226 &   -349.0040 \\
                & \pm  0.1794 & \pm  1.9523 & \pm  8.5410 & \pm 10.1673 \\
$t$             &      1.9977 &    -22.1227 &     70.8515 &    -50.7264 \\
                & \pm  0.2293 & \pm  2.6777 & \pm 10.1475 & \pm  7.8130 \\
$t \tau$        &     15.0237 &    -38.3450 &    183.8178 &   -160.4965 \\
                & \pm  0.3419 & \pm  1.8260 & \pm  5.2644 & \pm  3.7129 \\
$ls$            &     -2.6164 &     39.4240 &   -217.0569 &   -109.6725 \\
                & \pm  0.1947 & \pm  3.3849 & \pm 17.5511 & \pm 10.2746 \\
$ls \tau$       &      0.0069 &      2.5897 &    -26.5807 &    -77.5825 \\
                & \pm  0.0944 & \pm  1.1685 & \pm  5.5782 & \pm  3.3168 \\
$l2$            &      1.4358 &    -23.5937 &     67.8942 &    144.1521 \\
                & \pm  0.1809 & \pm  3.5108 & \pm 18.4785 & \pm 16.7585 \\
$l2 \tau$       &     -0.4106 &      8.3379 &    -82.9823 &    175.1091 \\
                & \pm  0.0950 & \pm  1.4331 & \pm  6.2147 & \pm  5.7715 \\
$l2 \sigma$     &     -0.0990 &      2.2549 &    -51.8708 &    175.0991 \\
                & \pm  0.1040 & \pm  1.5679 & \pm  6.6876 & \pm  6.2497 \\
$l2 \sigma \tau$&     -0.2667 &      6.6299 &    -55.3425 &    100.7191 \\
                & \pm  0.0343 & \pm  0.5087 & \pm  2.1657 & \pm  2.3042 \\
$ls2$           &      0.4583 &    -11.6586 &    150.5353 &   -302.1105 \\
                & \pm  0.2816 & \pm  4.9506 & \pm 22.8210 & \pm 17.1765 \\
$ls2 \tau$      &      0.7156 &    -18.8891 &    141.7216 &   -182.7536 \\
                & \pm  0.1273 & \pm  1.8340 & \pm  7.5529 & \pm  5.7410 \\
$T$             &      0.6379 &     -7.9042 &     24.2319 &    -19.7389 \\
                & \pm  0.1996 & \pm  2.6738 & \pm  9.9460 & \pm 10.6364 \\
$\sigma T$      &     -0.6379 &      7.9042 &    -24.2319 &     19.7389 \\
                & \pm  0.1996 & \pm  2.6738 & \pm  9.9460 & \pm 10.6364 \\
$l2 T$          &     -0.1063 &      1.3174 &     -4.0386 &      3.2898 \\
                & \pm  0.0333 & \pm  0.4456 & \pm  1.6577 & \pm  1.7727 \\
$l2 \sigma T$   &      0.1063 &     -1.3174 &      4.0386 &     -3.2898 \\
                & \pm  0.0333 & \pm  0.4456 & \pm  1.6577 & \pm  1.7727 \\
 \end{tabular*}
 \end{ruledtabular}
\end{table}

\begin{table*}[ht]
	\centering
	\caption{Deuteron static properties compared with empirical/recommended 
          values and high-quality potentials
          calculations. We list binding energy $E_d$, asymptotic D/S
          ratio $\eta$, asymptotic S-wave amplitude $A_S$, mean
          squared matter radius $r_m$, quadrupole moment $Q_D$ and D-wave
          probability $P_D$.}
	\label{tab:DeuteronP}
	\begin{tabular*}{\textwidth}{@{\extracolsep{\fill}}l l l l l l l l l}
      \hline
      \hline
            & This work & Emp./Rec.\cite{VanDerLeun1982261,Borbély198517,Rodning:1990zz,Klarsfeld1986373,Bishop:1979zz,deSwart:1995ui} & $\delta$-shell~\cite{Perez:2013mwa} & Nijm I~\cite{Stoks:1994wp}   & Nijm II~\cite{Stoks:1994wp}  & Reid93~\cite{Stoks:1994wp}   & AV18~\cite{Wiringa:1994wb} & CD-Bonn~\cite{Machleidt:2000ge}  \\
      \hline
		$E_d$(MeV)              & Input       & 2.224575(9)    & Input       & Input    & Input    & Input    & Input  & Input  \\
		$\eta$                  & 0.02448(5)  & 0.0256(5)      & 0.02493(8)  & 0.02534  & 0.02521  & 0.02514  & 0.0250 & 0.0256 \\
		$A_S ({\rm fm}^{1/2})$  & 0.8885(3)   & 0.8845(8)     & 0.8829(4)   & 0.8841   & 0.8845   & 0.8853   & 0.8850 & 0.8846 \\
		$r_m ({\rm fm})$        & 1.9744(6)   & 1.971(6)       & 1.9645(9)   & 1.9666   & 1.9675   & 1.9686   & 1.967  &  1.966 \\
		$Q_D ({\rm fm}^{2}) $   & 0.2645(7)   & 0.2859(3)      & 0.2679(9)   & 0.2719   & 0.2707   & 0.2703   & 0.270  & 0.270  \\
		$P_D$                   & 5.30(4)     & 5.67(4)        & 5.62(5)     & 5.664    & 5.635    & 5.699    & 5.76   & 4.85   \\
      \hline \hline
	\end{tabular*}
\end{table*}

\begin{table*}[htb]
 \footnotesize
 \caption{\label{tab:ppIsovectorPS} pp isovector phaseshifts.}
 \begin{ruledtabular}
 \begin{tabular*}{\textwidth}{@{\extracolsep{\fill}} r *{12}{D{.}{.}{3.3}}}
$E_{\rm LAB}$&\multicolumn{1}{c}{$^1S_0$}&\multicolumn{1}{c}{$^1D_2$}&\multicolumn{1}{c}{$^1G_4$}&\multicolumn{1}{c}{$^3P_0$}&\multicolumn{1}{c}{$^3P_1$}&\multicolumn{1}{c}{$^3F_3$}&\multicolumn{1}{c}{$^3P_2$}&\multicolumn{1}{c}{$\epsilon_2$}&\multicolumn{1}{c}{$^3F_2$}&\multicolumn{1}{c}{$^3F_4$}&\multicolumn{1}{c}{$\epsilon_4$}&\multicolumn{1}{c}{$^3H_4$}\\ 
  \hline 
  1 &    32.666 &     0.001 &     0.000 &     0.133 &    -0.080 &    -0.000 &     0.013 &    -0.001 &     0.000 &     0.000 &    -0.000 &     0.000\\
    & \pm 0.003 & \pm 0.000 & \pm 0.000 & \pm 0.000 & \pm 0.000 & \pm 0.000 & \pm 0.000 & \pm 0.000 & \pm 0.000 & \pm 0.000 & \pm 0.000 & \pm 0.000\\
  5 &    54.834 &     0.042 &     0.000 &     1.578 &    -0.899 &    -0.004 &     0.205 &    -0.052 &     0.002 &     0.000 &    -0.000 &     0.000\\
    & \pm 0.006 & \pm 0.000 & \pm 0.000 & \pm 0.002 & \pm 0.001 & \pm 0.000 & \pm 0.001 & \pm 0.000 & \pm 0.000 & \pm 0.000 & \pm 0.000 & \pm 0.000\\
 10 &    55.223 &     0.163 &     0.003 &     3.729 &    -2.053 &    -0.031 &     0.628 &    -0.201 &     0.013 &     0.001 &    -0.004 &     0.000\\
    & \pm 0.010 & \pm 0.000 & \pm 0.000 & \pm 0.005 & \pm 0.002 & \pm 0.000 & \pm 0.002 & \pm 0.000 & \pm 0.000 & \pm 0.000 & \pm 0.000 & \pm 0.000\\
 25 &    48.694 &     0.688 &     0.040 &     8.616 &    -4.892 &    -0.233 &     2.440 &    -0.815 &     0.103 &     0.018 &    -0.049 &     0.004\\
    & \pm 0.014 & \pm 0.001 & \pm 0.000 & \pm 0.016 & \pm 0.007 & \pm 0.000 & \pm 0.005 & \pm 0.001 & \pm 0.000 & \pm 0.000 & \pm 0.000 & \pm 0.000\\
 50 &    39.040 &     1.701 &     0.152 &    11.601 &    -8.186 &    -0.704 &     5.823 &    -1.735 &     0.328 &     0.099 &    -0.197 &     0.026\\
    & \pm 0.018 & \pm 0.003 & \pm 0.000 & \pm 0.030 & \pm 0.013 & \pm 0.001 & \pm 0.009 & \pm 0.003 & \pm 0.001 & \pm 0.001 & \pm 0.000 & \pm 0.000\\
100 &    25.452 &     3.820 &     0.414 &     9.567 &   -13.010 &    -1.546 &    11.074 &    -2.727 &     0.774 &     0.444 &    -0.553 &     0.107\\
    & \pm 0.034 & \pm 0.008 & \pm 0.001 & \pm 0.052 & \pm 0.017 & \pm 0.008 & \pm 0.013 & \pm 0.007 & \pm 0.007 & \pm 0.004 & \pm 0.001 & \pm 0.000\\
150 &    15.567 &     5.642 &     0.702 &     4.732 &   -17.296 &    -2.070 &    14.058 &    -2.980 &     1.132 &     0.991 &    -0.881 &     0.201\\
    & \pm 0.050 & \pm 0.014 & \pm 0.005 & \pm 0.064 & \pm 0.026 & \pm 0.019 & \pm 0.020 & \pm 0.010 & \pm 0.015 & \pm 0.009 & \pm 0.002 & \pm 0.002\\
200 &     7.490 &     7.058 &     1.032 &    -0.388 &   -21.412 &    -2.308 &    15.663 &    -2.875 &     1.337 &     1.642 &    -1.158 &     0.292\\
    & \pm 0.064 & \pm 0.022 & \pm 0.011 & \pm 0.064 & \pm 0.037 & \pm 0.031 & \pm 0.025 & \pm 0.017 & \pm 0.024 & \pm 0.014 & \pm 0.004 & \pm 0.005\\
250 &     0.500 &     8.276 &     1.385 &    -5.174 &   -25.335 &    -2.371 &    16.506 &    -2.603 &     1.289 &     2.272 &    -1.381 &     0.380\\
    & \pm 0.080 & \pm 0.026 & \pm 0.017 & \pm 0.066 & \pm 0.052 & \pm 0.044 & \pm 0.032 & \pm 0.023 & \pm 0.032 & \pm 0.019 & \pm 0.005 & \pm 0.011\\
300 &    -5.699 &     9.537 &     1.713 &    -9.460 &   -29.016 &    -2.385 &    16.892 &    -2.253 &     0.891 &     2.768 &    -1.556 &     0.478\\
    & \pm 0.102 & \pm 0.032 & \pm 0.022 & \pm 0.087 & \pm 0.073 & \pm 0.061 & \pm 0.044 & \pm 0.031 & \pm 0.041 & \pm 0.026 & \pm 0.006 & \pm 0.018\\
350 &   -11.239 &    10.974 &     1.959 &   -13.221 &   -32.431 &    -2.461 &    16.977 &    -1.875 &     0.091 &     3.056 &    -1.691 &     0.608\\
    & \pm 0.130 & \pm 0.059 & \pm 0.027 & \pm 0.124 & \pm 0.101 & \pm 0.084 & \pm 0.060 & \pm 0.042 & \pm 0.054 & \pm 0.045 & \pm 0.006 & \pm 0.025\\
 \end{tabular*}
 \end{ruledtabular}
\end{table*}

\begin{table*}[htb]
 \footnotesize
 \caption{\label{tab:npIsovectorPS}np isovector phaseshifts.}
 \begin{ruledtabular}
  \begin{tabular*}{\textwidth}{@{\extracolsep{\fill}} r *{12}{D{.}{.}{3.3}}}
$E_{\rm LAB}$&\multicolumn{1}{c}{$^1S_0$}&\multicolumn{1}{c}{$^1D_2$}&\multicolumn{1}{c}{$^1G_4$}&\multicolumn{1}{c}{$^3P_0$}&\multicolumn{1}{c}{$^3P_1$}&\multicolumn{1}{c}{$^3F_3$}&\multicolumn{1}{c}{$^3P_2$}&\multicolumn{1}{c}{$\epsilon_2$}&\multicolumn{1}{c}{$^3F_2$}&\multicolumn{1}{c}{$^3F_4$}&\multicolumn{1}{c}{$\epsilon_4$}&\multicolumn{1}{c}{$^3H_4$}\\ 
  \hline 
  1 &    62.074 &     0.001 &     0.000 &     0.180 &    -0.108 &    -0.000 &     0.021 &    -0.001 &     0.000 &     0.000 &    -0.000 &     0.000\\
    & \pm 0.018 & \pm 0.000 & \pm 0.000 & \pm 0.000 & \pm 0.000 & \pm 0.000 & \pm 0.000 & \pm 0.000 & \pm 0.000 & \pm 0.000 & \pm 0.000 & \pm 0.000\\
  5 &    63.652 &     0.040 &     0.000 &     1.653 &    -0.940 &    -0.004 &     0.248 &    -0.048 &     0.002 &     0.000 &    -0.000 &     0.000\\
    & \pm 0.045 & \pm 0.000 & \pm 0.000 & \pm 0.002 & \pm 0.001 & \pm 0.000 & \pm 0.001 & \pm 0.000 & \pm 0.000 & \pm 0.000 & \pm 0.000 & \pm 0.000\\
 10 &    60.004 &     0.154 &     0.002 &     3.747 &    -2.073 &    -0.026 &     0.705 &    -0.185 &     0.011 &     0.001 &    -0.003 &     0.000\\
    & \pm 0.065 & \pm 0.000 & \pm 0.000 & \pm 0.006 & \pm 0.003 & \pm 0.000 & \pm 0.002 & \pm 0.000 & \pm 0.000 & \pm 0.000 & \pm 0.000 & \pm 0.000\\
 25 &    51.043 &     0.669 &     0.032 &     8.506 &    -4.896 &    -0.201 &     2.586 &    -0.768 &     0.089 &     0.015 &    -0.039 &     0.003\\
    & \pm 0.107 & \pm 0.001 & \pm 0.000 & \pm 0.017 & \pm 0.007 & \pm 0.000 & \pm 0.005 & \pm 0.001 & \pm 0.000 & \pm 0.000 & \pm 0.000 & \pm 0.000\\
 50 &    40.920 &     1.701 &     0.131 &    11.433 &    -8.251 &    -0.634 &     6.025 &    -1.688 &     0.295 &     0.089 &    -0.169 &     0.020\\
    & \pm 0.167 & \pm 0.003 & \pm 0.001 & \pm 0.031 & \pm 0.013 & \pm 0.001 & \pm 0.009 & \pm 0.003 & \pm 0.001 & \pm 0.001 & \pm 0.000 & \pm 0.000\\
100 &    27.691 &     3.863 &     0.365 &     9.314 &   -13.211 &    -1.447 &    11.261 &    -2.747 &     0.724 &     0.428 &    -0.505 &     0.090\\
    & \pm 0.268 & \pm 0.008 & \pm 0.007 & \pm 0.053 & \pm 0.018 & \pm 0.008 & \pm 0.014 & \pm 0.007 & \pm 0.007 & \pm 0.004 & \pm 0.001 & \pm 0.000\\
150 &    18.146 &     5.697 &     0.594 &     4.380 &   -17.569 &    -1.977 &    14.170 &    -3.042 &     1.083 &     0.981 &    -0.834 &     0.176\\
    & \pm 0.313 & \pm 0.014 & \pm 0.027 & \pm 0.064 & \pm 0.027 & \pm 0.020 & \pm 0.020 & \pm 0.010 & \pm 0.016 & \pm 0.009 & \pm 0.002 & \pm 0.002\\
200 &    10.161 &     7.111 &     0.838 &    -0.809 &   -21.717 &    -2.236 &    15.705 &    -2.938 &     1.295 &     1.643 &    -1.124 &     0.261\\
    & \pm 0.309 & \pm 0.022 & \pm 0.056 & \pm 0.064 & \pm 0.038 & \pm 0.032 & \pm 0.025 & \pm 0.017 & \pm 0.024 & \pm 0.014 & \pm 0.004 & \pm 0.005\\
250 &     3.068 &     8.331 &     1.118 &    -5.626 &   -25.658 &    -2.322 &    16.495 &    -2.644 &     1.248 &     2.280 &    -1.369 &     0.347\\
    & \pm 0.304 & \pm 0.026 & \pm 0.085 & \pm 0.067 & \pm 0.053 & \pm 0.045 & \pm 0.032 & \pm 0.024 & \pm 0.032 & \pm 0.019 & \pm 0.005 & \pm 0.011\\
300 &    -3.345 &     9.601 &     1.434 &    -9.916 &   -29.352 &    -2.356 &    16.840 &    -2.271 &     0.841 &     2.775 &    -1.566 &     0.448\\
    & \pm 0.345 & \pm 0.033 & \pm 0.102 & \pm 0.089 & \pm 0.074 & \pm 0.062 & \pm 0.045 & \pm 0.031 & \pm 0.042 & \pm 0.026 & \pm 0.006 & \pm 0.018\\
350 &    -9.144 &    11.052 &     1.763 &   -13.666 &   -32.782 &    -2.447 &    16.891 &    -1.879 &     0.022 &     3.053 &    -1.720 &     0.583\\
    & \pm 0.441 & \pm 0.062 & \pm 0.105 & \pm 0.127 & \pm 0.103 & \pm 0.085 & \pm 0.061 & \pm 0.043 & \pm 0.055 & \pm 0.047 & \pm 0.006 & \pm 0.025\\
 \end{tabular*}
 \end{ruledtabular}
\end{table*}

\begin{table*}
 \footnotesize
 \caption{\label{tab:npIsoscalarPS}np isoscalar phaseshifts.}
 \begin{ruledtabular}
 \begin{tabular*}{\textwidth}{@{\extracolsep{\fill}} r *{12}{D{.}{.}{3.3}}}
 $E_{\rm LAB}$&\multicolumn{1}{c}{$^1P_1$}&\multicolumn{1}{c}{$^1F_3$}&\multicolumn{1}{c}{$^3D_2$}&\multicolumn{1}{c}{$^3G_4$}&\multicolumn{1}{c}{$^3S_1$}&\multicolumn{1}{c}{$\epsilon_1$}&\multicolumn{1}{c}{$^3D_1$}&\multicolumn{1}{c}{$^3D_3$}&\multicolumn{1}{c}{$\epsilon_3$}&\multicolumn{1}{c}{$^3G_3$}\\ 
  \hline 
  1 &    -0.186 &    -0.000 &     0.006 &     0.000 &   147.624 &     0.102 &    -0.005 &     0.000 &     0.000 &    -0.000\\
    & \pm 0.000 & \pm 0.000 & \pm 0.000 & \pm 0.000 & \pm 0.009 & \pm 0.000 & \pm 0.000 & \pm 0.000 & \pm 0.000 & \pm 0.000\\
  5 &    -1.493 &    -0.010 &     0.218 &     0.001 &   117.905 &     0.638 &    -0.177 &     0.002 &     0.012 &    -0.000\\
    & \pm 0.004 & \pm 0.000 & \pm 0.000 & \pm 0.000 & \pm 0.020 & \pm 0.003 & \pm 0.000 & \pm 0.000 & \pm 0.000 & \pm 0.000\\
 10 &    -3.058 &    -0.064 &     0.843 &     0.012 &   102.230 &     1.086 &    -0.661 &     0.007 &     0.080 &    -0.003\\
    & \pm 0.010 & \pm 0.000 & \pm 0.001 & \pm 0.000 & \pm 0.028 & \pm 0.007 & \pm 0.001 & \pm 0.000 & \pm 0.000 & \pm 0.000\\
 25 &    -6.337 &    -0.421 &     3.698 &     0.170 &    80.068 &     1.653 &    -2.735 &     0.058 &     0.552 &    -0.053\\
    & \pm 0.034 & \pm 0.000 & \pm 0.005 & \pm 0.000 & \pm 0.041 & \pm 0.018 & \pm 0.005 & \pm 0.003 & \pm 0.000 & \pm 0.000\\
 50 &    -9.603 &    -1.143 &     8.951 &     0.722 &    62.105 &     1.955 &    -6.276 &     0.376 &     1.609 &    -0.264\\
    & \pm 0.071 & \pm 0.003 & \pm 0.020 & \pm 0.000 & \pm 0.053 & \pm 0.035 & \pm 0.013 & \pm 0.013 & \pm 0.002 & \pm 0.000\\
100 &   -14.089 &    -2.291 &    17.299 &     2.181 &    42.633 &     2.428 &   -11.922 &     1.599 &     3.451 &    -0.989\\
    & \pm 0.113 & \pm 0.022 & \pm 0.049 & \pm 0.005 & \pm 0.065 & \pm 0.066 & \pm 0.030 & \pm 0.038 & \pm 0.011 & \pm 0.004\\
150 &   -17.844 &    -3.102 &    22.164 &     3.665 &    30.269 &     2.980 &   -16.143 &     2.830 &     4.700 &    -1.898\\
    & \pm 0.129 & \pm 0.052 & \pm 0.060 & \pm 0.019 & \pm 0.066 & \pm 0.085 & \pm 0.045 & \pm 0.054 & \pm 0.024 & \pm 0.013\\
200 &   -21.036 &    -3.775 &    24.449 &     5.065 &    20.890 &     3.517 &   -19.526 &     3.690 &     5.536 &    -2.851\\
    & \pm 0.148 & \pm 0.080 & \pm 0.073 & \pm 0.041 & \pm 0.067 & \pm 0.093 & \pm 0.059 & \pm 0.061 & \pm 0.034 & \pm 0.029\\
250 &   -23.623 &    -4.421 &    25.137 &     6.379 &    13.208 &     4.007 &   -22.339 &     4.222 &     6.150 &    -3.787\\
    & \pm 0.181 & \pm 0.100 & \pm 0.096 & \pm 0.066 & \pm 0.088 & \pm 0.099 & \pm 0.072 & \pm 0.074 & \pm 0.039 & \pm 0.048\\
300 &   -25.653 &    -5.078 &    24.920 &     7.604 &     6.681 &     4.476 &   -24.681 &     4.578 &     6.648 &    -4.692\\
    & \pm 0.222 & \pm 0.116 & \pm 0.121 & \pm 0.086 & \pm 0.131 & \pm 0.114 & \pm 0.088 & \pm 0.099 & \pm 0.047 & \pm 0.067\\
350 &   -27.236 &    -5.734 &    24.242 &     8.712 &     1.036 &     4.956 &   -26.586 &     4.876 &     7.067 &    -5.568\\
    & \pm 0.266 & \pm 0.145 & \pm 0.147 & \pm 0.097 & \pm 0.183 & \pm 0.137 & \pm 0.107 & \pm 0.130 & \pm 0.065 & \pm 0.082\\
 \end{tabular*}
 \end{ruledtabular}
\end{table*}

\section{The OPE-Gaussian potential}
\label{sec:OPEgauss}

In the present section we provide a rather simple local form of the
potential Eq.~(\ref{eq:potential}) and Eq.~(\ref{eq:potential-short})
based on Gaussian functions
\begin{eqnarray}
F_{i,n}(r) = e^{- r^2/(2a_i^2)} 
\end{eqnarray}
where we have taken the parameters as $a_i = a/(1+i) $. The parameter
$a$ is used as a fitting variable. With this potential we get
$\chi^2/\nu = 1.06 $. The resulting 42 fitting parameters (41
independent partial wave coefficients $(V_i)^{JS}_{l,l'} $ and the
Gaussian width $a$ are listed with their uncertainties in
Table~\ref{tab:FitParameters}. The $V_{i,n}$ operator coefficients are
given in Table~\ref{tab:OperatorParameters}~\footnote{The many digits
  are provided to guarantee numerical reproducibility of results,
  since we find strong correlations among the parameters. We thank
  Eduardo Garrido numerical checks.}. The linear transformation from
partial wave coefficients $(V_i)^{JS}_{l,l'} $ to the $V_{i,n}$
operator coefficients has been given explicitly in
Ref.~\cite{Perez:2013jpa}.  In Fig.~\ref{fig:corr} we depict the
correlation matrix, Eq.~(\ref{eq:correlation}) for the partial wave
parameters listed in Table~\ref{tab:FitParameters}, where a similar
correlation pattern to the OPE-DS one is observed. Deuteron properties
for this potential compared with calculations using other potentials
and empirical or recommended values can be looked up in
Table~\ref{tab:DeuteronP}.

The rotated QQ-plot of the scaled residuals for the OPE-Gaussian fit
to the $3\sigma$ self-consistent database can be seen in
Fig.~\ref{Fig:QQconfidence}. As we can see the TS test is passed
satisfactorily. On a more quantitative level we show on
Table~\ref{tab:Cmoments} the moments test.  The resulting p-value of
the different normality tests are given in
Tables~\ref{tab:PearsontestResults}, \ref{tab:KStestResults} and
\ref{tab:TStestResults} for the Pearson, KS and TS tests
respectively. As we see all tests are satisfactorily passed except for
the TS where a tiny scaling of the residuals by a Birge factor of
$\sqrt{\chi^2/\nu} = 1.03$, corresponding to a global enlargement of
the provided experimental errors by $3\%$, allows to restore
normality.  Thus, we are entitled to propagate the uncertainties of
the data to derived quantities through the determined parameters
$V_{i,n}$ with errors and their corresponding correlations, see
Eq.~(\ref{eq:error-prop}).

In Figs.~\ref{fig:local-pot} and \ref{fig:operator-pot} we show the
OPE-Gaussian potential in partial wave and operator basis respectively
with the error bands propagated with the corresponding correlation
matrix from the fit to the experimental data. As we see, these error
bands are smaller than the discrepancy with Reid93~\cite{Stoks:1994wp}
NijmII~\cite{Stoks:1994wp} and AV18~\cite{Wiringa:1994wb}.  This may
be a hint that systematic errors induced by the bias involved in the
choice of the several potentials, as first noted in
Ref.~\cite{NavarroPerez:2012vr,Perez:2012kt,Amaro:2013zka}, may indeed
play a relevant role in the total evaluation of nuclear uncertainties.

\begin{figure*}[h]
\begin{center}
\epsfig{figure=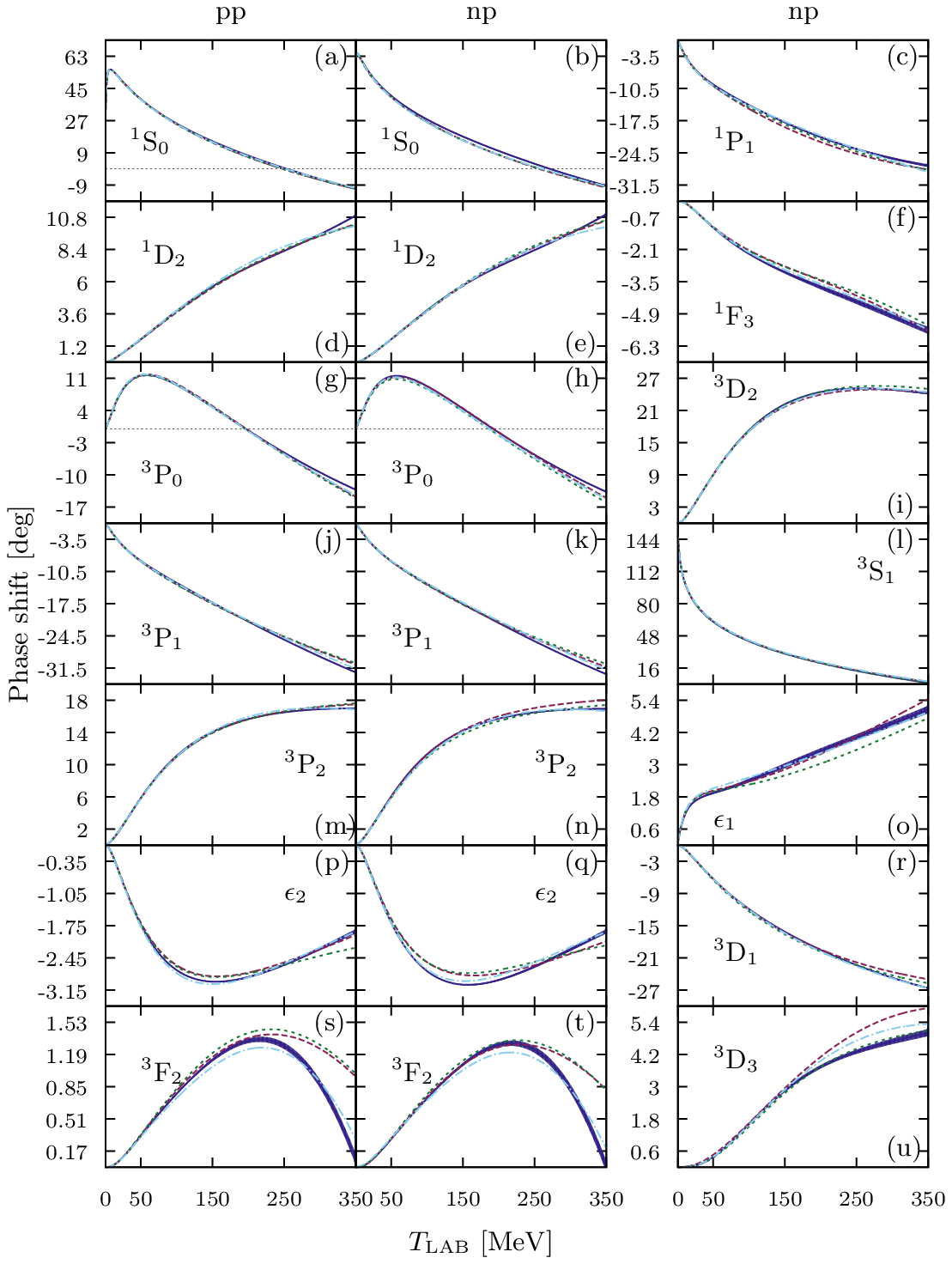,height=22cm,width=14.7cm}
\end{center}
\caption{(Color online) Lowest np and pp phase shifts (in degrees) and their errors
for the present OPE+Gaussian analysis (blue band)
  Reid93~\cite{Stoks:1994wp} (red dashed) NijmII~\cite{Stoks:1994wp}
  (green dotted) and AV18~\cite{Wiringa:1994wb} (light-blue
  dashed-dotted) as a function of the LAB energy
  (in MeV).}
\label{fig:all-phases}
\end{figure*}

In Fig.~\ref{fig:all-phases} we present the lowest np and pp phase
shifts and their errors based on the OPE-Gaussian potential and
compared with the Reid93~\cite{Stoks:1994wp},
NijmII~\cite{Stoks:1994wp} and AV18~\cite{Wiringa:1994wb} potential
phases. In Tables~\ref{tab:ppIsovectorPS} , \ref{tab:npIsovectorPS}
and \ref{tab:npIsoscalarPS} the low angular momentum phases as a
function of the LAB energy with their errors propagated from the fit
are listed.

\begin{figure*}[h]
\begin{center}
\epsfig{figure=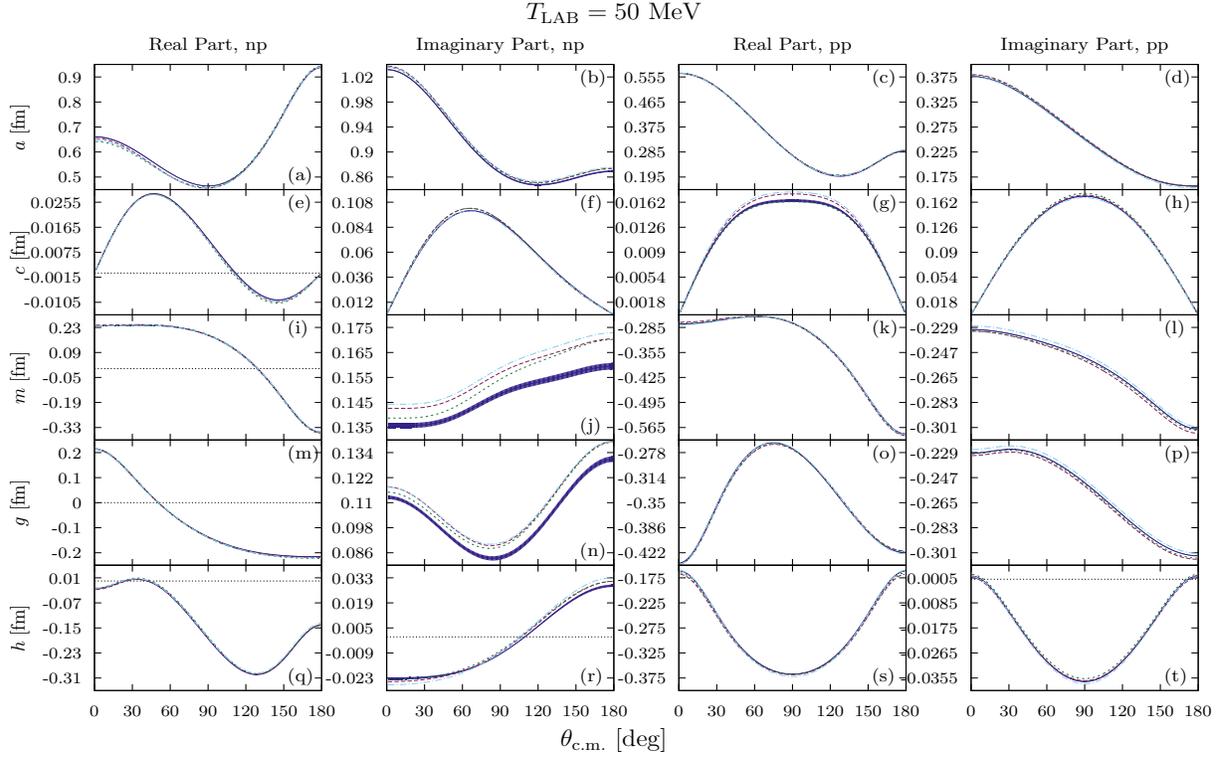,height=10cm,width=16cm}
\end{center}
\caption{(Color on-line) np (left) and pp (right) Wolfenstein
  parameters (in fm) as a function of the CM angle (in degrees) and
  for $E_{\rm LAB}=50 {\rm MeV}$. We compare our fit (blue band) with 
  Reid93~\cite{Stoks:1994wp} (red dashed) NijmII~\cite{Stoks:1994wp}
  (green dotted) and AV18~\cite{Wiringa:1994wb} (light-blue
  dashed-dotted).}
\label{FigWolfenstein050}
\end{figure*}

\begin{figure*}[h]
\begin{center}
\epsfig{figure=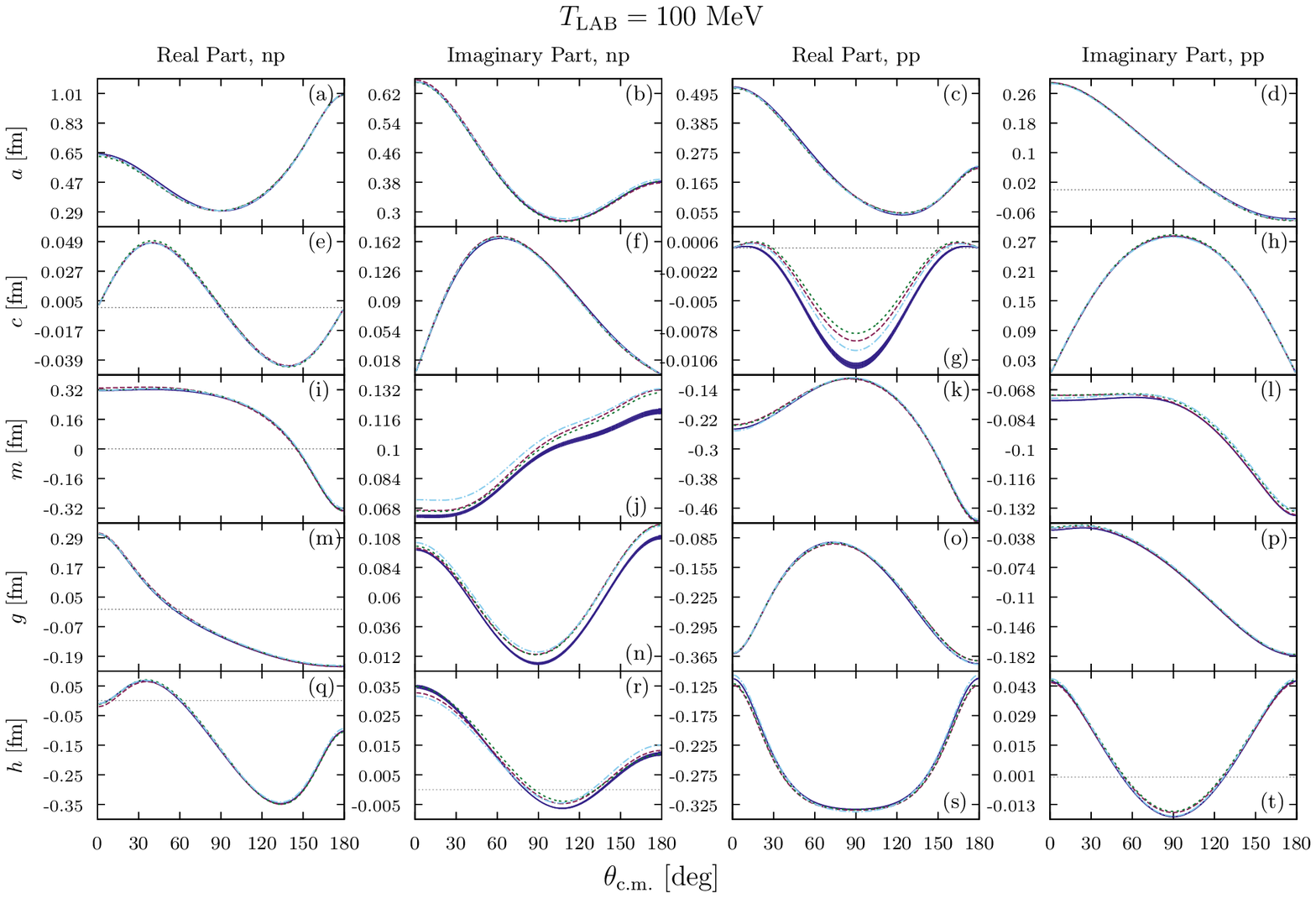,height=10cm,width=16cm}
\end{center}
\caption{(Color on-line) Same as in Fig.~\ref{FigWolfenstein050} but for  $E_{\rm LAB}=100 {\rm
    MeV}$.}
\label{FigWolfenstein100}
\end{figure*}

\begin{figure*}[h]
\begin{center}
\epsfig{figure=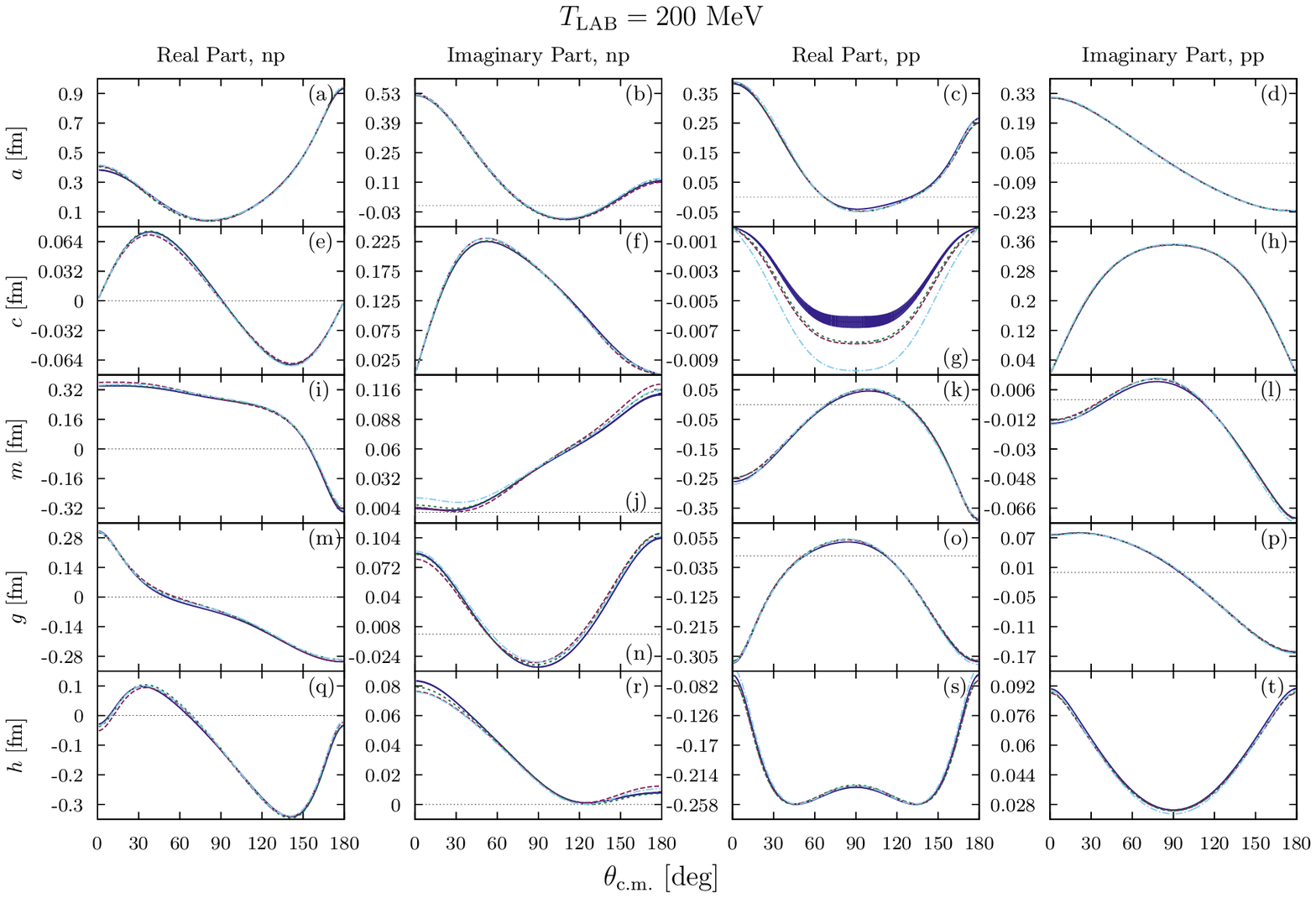,height=10cm,width=16cm}
\end{center}
\caption{(Color on-line) Same as in Fig.~\ref{FigWolfenstein050} but for  $E_{\rm LAB}=200 {\rm
    MeV}$.}
\label{FigWolfenstein200}
\end{figure*}

\begin{figure*}[h]
\begin{center}
\epsfig{figure=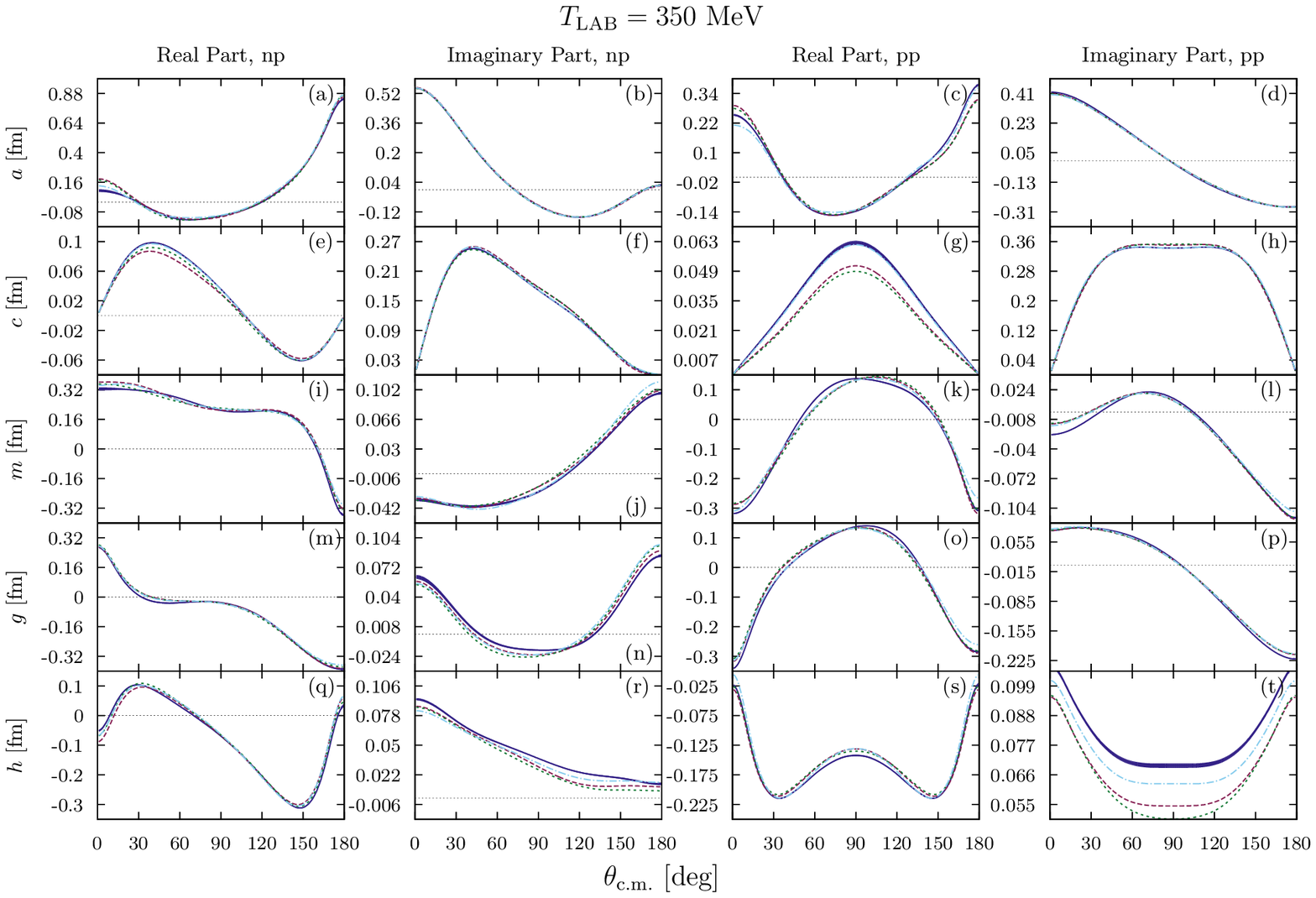,height=10cm,width=16cm}
\end{center}
\caption{(Color on-line) Same as in Fig.~\ref{FigWolfenstein050} but for  $E_{\rm LAB}=350 {\rm
    MeV}$.}
\label{FigWolfenstein350}
\end{figure*}

The resulting Wolfenstein parameters, Eq.~(\ref{eq:wolfenstein}), for
the OPE-Gaussian potential are depicted in
Figs.~\ref{FigWolfenstein050}, \ref{FigWolfenstein100},
\ref{FigWolfenstein200} and \ref{FigWolfenstein350} for LAB energies
$50,100,200$ and $350$ MeV respectively with their corresponding errors.
For comparison we also show the same quantities calculated with the
1993-high quality Reid93~\cite{Stoks:1994wp},
NijmII~\cite{Stoks:1994wp} and AV18~\cite{Wiringa:1994wb} potentials.

\section{Conclusions and outlook}
\label{sec:conclusions}

We summarize our main points. The determination of uncertainties in
theoretical nuclear physics is one of the most urgent issues to be
solved in order to establish the predictive power of {\it ab initio}
nuclear structure calculations. One certain source for these
uncertainties is the errors of the phenomenological NN interaction
stemming from the finite accuracy of experimental scattering data as
well as local scarcity in certain regions of the $(\theta,E)$ plane
and an abundance bias in some other regions.  Any statistical analysis
of this sort {\it assumes} a model both for the signal and the noise
which can only be checked {\it a posteriori}. In order to carry out
such an analysis the lack of bias in the data and the model has to be
established with a given confidence level. If normal errors on the
data are assumed, the check can be made by applying normality tests to
the residuals between the fitted model and the experimental data. We
have used some classical tests and the highly demanding recently
proposed Tail-sensitive Quantile-quantile test with a confidence level
of $95\%$. Based on the outcome there is no serious reason to doubt on
the normality of residuals of the $3\sigma$ self-consistent database
obtained in our PWA of np and pp scattering data below pion production
threshold.

We remind that this normality test actually checks for the assumption,
underlying any least squares $\chi^2$ fit, that the data themselves
follow a normal distribution. With this fixed database one can then look
for different representations of the potential which facilitate a
straightforward implementation in any of the many available powerful
methods which are currently available for solving the multi-nucleon
problem.

We provide a user friendly potential which consists of a short range
local part with 21-operators multiplying a linear superposition of
Gaussian functions. The resulting fitted potential passes the
normality tests satisfactorily and hence can be used to estimate
statistical uncertainties stemming from NN scattering data.

Our findings here seem to confirm a previous study of us when we
compare the current OPE-Gauss potential including {\it statistical
  error bands} with previous potentials such as NijmII, Red93 or AV18
(without statistical bands); errors in the potential are dominated by
the form of the potential, rather than by the experimental
data. Nonetheless, a thorough study of these kind of errors requires
repeating the present analysis with an identical database with the
{\it most general} potentials and functional forms, and looking for
discrepancies in the nuclear structure calculations outcome.

\vskip0.5cm 

{\it We thank El\'{\i}as Moreno for a statistician's point of view,
  Antonio Bueno for disclosing the experimentalist's feelings and
  Lorenzo Luis Salcedo for an introduction to the Bayesian
  approach. We also thank Eduardo Garrido for numerical checks.}


\end{document}